\documentclass[]{aa}
\usepackage{epsfig,lscape}
\begin{document}
\def\lsim{\vcenter{\hbox{$<$}\offinterlineskip\hbox{$\sim$}}}
\def\gsim{\vcenter{\hbox{$>$}\offinterlineskip\hbox{$\sim$}}}
\thesaurus{06(08.03.1; 08.03.4; 08.13.2; 08.16.4; 11.13.1; 13.09.6)}
\title{ISO observations of obscured Asymptotic Giant Branch stars in the Large
       Magellanic Cloud\thanks{This paper is based on observations with the
       Infrared Space Observatory (ISO). ISO is an ESA project with
       instruments funded by ESA member states (especially the PI countries:
       France, Germany, The Netherlands and the United Kingdom) and with the
       participation of ISAS and NASA.}}
\author{Norman R. Trams\inst{1}, Jacco Th. van Loon\inst{2}, L.B.F.M.
        Waters\inst{2,3}, Albert A. Zijlstra\inst{4}, Cecile Loup\inst{5},
        Patricia A. Whitelock\inst{6}, M.A.T. Groenewegen\inst{7},
        Joris A.D.L. Blommaert\inst{8}, Ralf Siebenmorgen\inst{8}, A.
        Heske\inst{8} \and Michael W. Feast\inst{9}}
\institute{Integral Science Operations Centre, Astrophysics Div., Science
           Dep., ESTEC, P.O.Box 299, NL-2200 AG Noordwijk, The Netherlands
      \and Astronomical Institute, University of Amsterdam, Kruislaan 403,
           NL-1098 SJ Amsterdam, The Netherlands
      \and Space Research Organization Netherlands, Landleven 12, NL-9700 AV
           Groningen, The Netherlands
      \and University of Manchester Institute of Science and Technology,
           P.O.Box 88, Manchester M60 1QD, United Kingdom
      \and Institut d'Astrophysique de Paris, 98bis Boulevard Arago, F-75014
           Paris, France
      \and South African Astronomical Observatory, P.O.Box 9, 7935
           Observatory, South Africa
      \and Max-Planck Institut f\"{u}r Astrophysik, Karl-Schwarzschild
           Stra{\ss}e 1, D-85740 Garching bei M\"{u}nchen, Germany
      \and ISO Data Centre, Astrophysics Division, Science Department of ESA,
           Villafranca del Castillo, P.O.Box 50727, E-28080 Madrid, Spain
      \and Astronomy Department, University of Cape Town, 7700 Rondebosch,
           South Africa}
\offprints{N.R.\ Trams, ntrams@astro.estec.esa.nl}
\date{Received date; accepted date}
\maketitle
\markboth{Trams et al.: ISO observations of AGB stars in LMC}
         {Trams et al.: ISO observations of AGB stars in LMC}
\begin{abstract}

We present ISO photometric and spectroscopic observations of a sample of 57
bright Asymptotic Giant Branch stars and red supergiants in the Large
Magellanic Cloud, selected on the basis of IRAS colours indicative of high
mass-loss rates. PHOT-P and PHOT-C photometry at 12, 25 and 60 $\mu$m and CAM
photometry at 12 $\mu$m are used in combination with quasi-simultaneous
ground-based near-IR photometry to construct colour-colour diagrams for all
stars in our sample. PHOT-S and CAM-CVF spectra in the 3 to 14 $\mu$m region
are presented for 23 stars. From the colour-colour diagrams and the spectra,
we establish the chemical types of the dust around 49 stars in this sample.
Many stars have carbon-rich dust. The most luminous carbon star in the
Magellanic Clouds has also a (minor) oxygen-rich component. OH/IR stars have
silicate absorption with emission wings. The unique dataset presented here
allows a detailed study of a representative sample of thermal-pulsing AGB
stars with well-determined luminosities.

\end{abstract}

\section{Introduction}

One of the least expected achievements of the Infra-Red Astronomical Satellite
(IRAS; Neugebauer et al.\ 1984) was the detection of a large number of mid-IR
point sources in the Large Magellanic Cloud (LMC) just above its limits of
sensitivity (IRAS Point Source Catalogue; Schwering \& Israel 1990). Many of
these are candidates for intermediate-mass stars at the tip of the Asymptotic
Giant Branch (AGB). Their lives drawing to a close, these stars are shedding
their stellar mantles at rates of up to $10^{-4}$ M$_\odot$ yr$^{-1}$. Their
dusty circumstellar envelopes (CSEs) obscure the optical light from the star
and become very bright IR objects. The details of the evolution and mass loss
of AGB stars are poorly understood. The study of galactic samples of AGB stars
is severely hampered by the difficulty to determine accurate distances to
stars in the Milky Way. The distance to the LMC, however, is well known and
hence luminosities and mass-loss rates of AGB stars in the LMC may be
determined with a high degree of accuracy.

Early explorations of the IRAS data in combination with ground-based near-IR
observations resulted in the first identifications of mid-IR sources in the
LMC with obscured AGB stars (Reid et al.\ 1990; Wood et al.\ 1992). We have
successfully increased the sample of known AGB counterparts of IRAS sources in
the LMC from a dozen to more than 50 stars (Loup et al.\ 1997; Zijlstra et
al.\ 1996; van Loon et al.\ 1997, 1998a: Papers I to IV). We attempted to
classify their photospheres and CSEs as oxygen- or carbon-dominated, but for
the majority of the stars this could not be done conclusively. There remained
therefore considerable uncertainty about the luminosity distribution of the
obscured carbon stars. This information is important for testing current
understanding of the evolution of AGB stars, including dredge-up of carbon and
nuclear burning at the bottom of the convective mantle (Hot Bottom Burning,
HBB).

%
% TABLE 1
%
\begin{table*}
\caption[]{IRAS detected stars observed with ISO: names (LI stands for LI-LMC
(Schwering \& Israel 1990), TRM is from Reid et al.\ (1990), HV is from
Payne-Gaposchkin (1971), SP is from Sanduleak \& Philip (1977) and WOH is from
Westerlund et al.\ (1981); sources will be referenced hereafter by their
bold-faced names), ISO pointing coordinates (J2000), and references:
1:  Hodge \& Wright (1969);
2:  Eggen (1971);
3:  Wright \& Hodge (1971);
4:  Dachs (1972);
5:  Sandage \& Tammann (1974);
6:  Glass (1979);
7:  Humphreys (1979);
8:  Blanco et al.\ (1980);
9:  Feast et al.\ (1980);
10: Bessell \& Wood (1983);
11: Wood et al.\ (1983);
12: Rebeirot et al.\ (1983);
13: Prevot et al.\ (1985);
14: Elias et al.\ (1985);
15: Wood et al.\ (1985);
16: Elias et al.\ (1986);
17: Wood et al.\ (1986);
18: Reid et al.\ (1988);
19: Reid (1989);
20: Hughes (1989);
21: Hughes \& Wood (1990);
22: Reid et al.\ (1990);
23: Hughes et al.\ (1991);
24: Wood et al.\ (1992);
25: Roche et al.\ (1993);
26: Groenewegen et al.\ (1995);
27: Zijlstra et al.\ (1996);
28: Ritossa et al.\ (1996);
29: van Loon et al.\ (1996);
30: van Loon et al.\ (1997);
31: Loup et al.\ (1997);
32: Oestreicher (1997);
33: van Loon et al.\ (1998a);
34: Groenewegen \& Blommaert (1998);
35: van Loon et al.\ (1998b);
36: van Loon et al.\ (1999)}
\begin{flushleft}
\begin{tabular}{llllllll}
\hline\hline
LI & IRAS               & TRM      & HV          &
 RA (2000)  & Decl (2000) &
 Other names                         & References                       \\
\hline
\multicolumn{8}{c}{{\em IRAS detected stars}} \\
\hline
1825   & {\bf 04286$-$6937} & --       & --          &
 04 28 30.3 & $-$69 30 49 &
 --                                  &                         27,31,33 \\
1844   & {\bf 04374$-$6831} & --       & --          &
 04 37 22.8 & $-$68 25 03 &
 --                                  &                         27,31,33 \\
4      & {\bf 04407$-$7000} & --       & --          &
 04 40 28.4 & $-$69 55 13 &
 --                                  &                         27,31,33 \\
57     & {\bf 04496$-$6958} & --       & --          &
 04 49 18.6 & $-$69 53 14 &
 --                                  &                   27,31,33,34,36 \\
60     & {\bf 04498$-$6842} & --       & --          &
 04 49 41.4 & $-$68 37 50 &
 --                                  &                      27,31,33,36 \\
77     & {\bf 04509$-$6922} & --       & --          &
 04 50 40.2 & $-$69 17 33 &
 --                                  &                   24,27,28,33,36 \\
92     & {\bf 04516$-$6902} & --       & --          &
 04 51 28.4 & $-$68 57 53 &
 --                                  &                         24,27,33 \\
121    & {\bf 04530$-$6916} & --       & --          &
 04 52 45.3 & $-$69 11 53 &
 --                                  &                         24,27,28 \\
141    & {\bf 04539$-$6821} & --       & --          &
 04 53 46.3 & $-$68 16 12 &
 --                                  &                         27,31,33 \\
153    &      04544$-$6849  & --       & --          &
 04 54 14.4 & $-$68 44 13 &
 {\bf SP77 30-6}, WOH SG66           &                12,13,20,21,27,31 \\
159    & {\bf 04545$-$7000} & --       & --          &
 04 54 09.8 & $-$69 56 00 &
 --                                  &                            24,27 \\
181    &      04553$-$6825  & --       & --          &
 04 55 10.1 & $-$68 20 35 &
 {\bf WOH G64}                       & 16,17,19,24,25,27,29,31,33,35,36 \\
198    & {\bf 04557$-$6753} & --       & --          &
 04 55 38.9 & $-$67 49 10 &
 --                                  &                         27,31,33 \\
203    &      04559$-$6931  & --       & {\bf 12501} &
 04 55 41.6 & $-$69 26 25 &
 SP77 31-20, WOH SG097               &          11,12,13,20,22,27,32,33 \\
297    & {\bf 05003$-$6712} & --       & --          &
 05 00 18.9 & $-$67 08 02 &
 --                                  &                   27,30,31,33,36 \\
310    & {\bf 05009$-$6616} & --       & --          &
 05 01 03.8 & $-$66 12 40 &
 --                                  &                      27,31,33,36 \\
383    &      05042$-$6720  &      48  &   {\bf 888} &
 05 04 14.3 & $-$67 16 17 &
 SP77 29-33, WOH SG140               &          5,6,7,11,14,18,22,31,32 \\
570    & {\bf 05112$-$6755} &       4  & --          &
 05 11 10.1 & $-$67 52 17 &
 --                                  &                   22,27,31,33,36 \\
571    & {\bf 05113$-$6739} &      24  & --          &
 05 11 13.7 & $-$67 36 35 &
 --                                  &                      22,27,31,33 \\
578    & --                 & {\bf 72} & --          &
 05 11 41.2 & $-$66 51 12 &
 --                                  &                      22,27,31,33 \\
612    &      05128$-$6728  &      43  &  {\bf 2360} &
 05 12 46.4 & $-$67 19 37 &
 SP77 37-24, WOH SG193               &           3,5,6,7,11,14,18,22,31 \\
1880   & {\bf 05128$-$6455} & --       & --          &
 05 13 04.6 & $-$64 51 40 &
 --                                  &                      27,31,33,36 \\
663    &      05148$-$6730  &      36  &   {\bf 916} &
 05 14 49.9 & $-$67 27 19 &
 SP77 37-35, WOH SG204               &         1,2,4,6,7,11,18,22,31,32 \\
793    & {\bf 05190$-$6748} &      20  & --          &
 05 18 56.7 & $-$67 45 06 &
 --                                  &                      22,27,31,33 \\
--     & --                 & {\bf 88} & --          &
 05 20 20.9 & $-$66 36 00 &
 --                                  &                22,27,31,33,34,36 \\
--     & --                 & {\bf 45} & --          &
 05 28 16.3 & $-$67 20 55 &
 --                                  &                      22,27,31,33 \\
1157   & {\bf 05295$-$7121} & --       & --          &
 05 28 40.8 & $-$71 19 13 &
 --                                  &                            27,31 \\
1130   & {\bf 05289$-$6617} &      99  & --          &
 05 29 02.6 & $-$66 15 31 &
 --                                  &                         22,27,31 \\
1145   & --                 & --       &  {\bf 5870} &
 05 29 03.5 & $-$69 06 47 &
 SP77 47-17, WOH SG331               &                       9,11,20,31 \\
1153   & {\bf 05294$-$7104} & --       & --          &
 05 28 47.8 & $-$71 02 29 &
 --                                  &                         24,27,31 \\
1164   & {\bf 05298$-$6957} & --       & --          &
 05 29 24.5 & $-$69 55 14 &
 --                                  &                      24,27,31,36 \\
1177   & {\bf 05300$-$6651} &      79  & --          &
 05 30 04.2 & $-$66 49 23 &
 --                                  &                      22,27,31,36 \\
1238   &      05316$-$6604  &     101  & --          &
 05 31 45.9 & $-$66 03 51 &
 {\bf WOH SG374}                     &                         22,27,31 \\
1281   &      05327$-$6757  &       5  &   {\bf 996} &
 05 32 36.0 & $-$67 55 08 &
 SP77 46-59, WOH SG388               &                 7,11,17,18,22,31 \\
1286   & {\bf 05329$-$6708} &      60  & --          &
 05 32 52.5 & $-$67 06 25 &
 --                                  &          17,22,24,26,27,31,33,36 \\
1345   & {\bf 05348$-$7024} & --       & --          &
 05 34 16.1 & $-$70 22 53 &
 --                                  &                         27,31,33 \\
1382   & {\bf 05360$-$6648} &      77  & --          &
 05 36 03.3 & $-$66 46 47 &
 --                                  &                      22,27,31,33 \\
1506   & {\bf 05402$-$6956} & --       & --          &
 05 39 44.6 & $-$69 55 18 &
 --                                  &                            24,27 \\
1756   & {\bf 05506$-$7053} & --       & --          &
 05 50 09.1 & $-$70 53 12 &
 --                                  &                         27,31,33 \\
1790   & {\bf 05558$-$7000} & --       & --          &
 05 55 20.8 & $-$70 00 05 &
 --                                  &                            27,31 \\
1795   & {\bf 05568$-$6753} & --       & --          &
 05 56 38.7 & $-$67 53 39 &
 --                                  &                            27,31 \\
\hline
\end{tabular}
\end{flushleft}
\normalsize
\end{table*}

57 obscured AGB stars and a few red supergiants (RSGs) in the LMC were
selected for Guaranteed Time and follow-up Open Time observations with the
Infrared Space Observatory (ISO; Kessler et al.\ 1996). The goals were to
obtain photometry at 12, 25 and 60 $\mu$m and to spectroscopically determine
the chemical types of the CSEs. The photometry, which covers the entire
spectral energy distributions (SEDs), can be modelled and used to derive
accurate luminosities and mass-loss rates. In this paper we present the ISO
data and classify sources as oxygen- or carbon-rich.

\section{Source selection}

%
% TABLE 2
%
\begin{table*}
\caption[]{The list of program stars without IRAS counterpart. The references
are as in Table 1. SHV is from Hughes (1989), BMB is from Blanco et al.\
(1980), WBP is from Wood et al.\ (1985) and GRV is from Reid et al.\ (1988).}
\begin{flushleft}
\begin{tabular}{llllllll}
\hline\hline
LI & IRAS               & TRM      & HV          &
 RA (2000)  & Decl (2000) &
 Other names                         & References                       \\
\hline
\multicolumn{8}{c}{{\em non-IRAS sources classified as C stars}} \\
\hline
--     & --                 & --       & --          &
 04 53 59.7 & $-$67 45 47 &
 {\bf SHV0454030$-$675031}           &                            20,21 \\
--     & --                 & --       & --          &
 05 02 28.7 & $-$69 20 10 &
 {\bf SHV0502469$-$692418}           &                         20,21,23 \\
--     & --                 & --       &  {\bf 2379} &
 05 14 46.3 & $-$67 55 47 &
 --                                  &                       3,10,11,20 \\
--     & --                 & --       & --          &
 05 20 46.8 & $-$69 01 25 &
 {\bf SHV0521050$-$690415}, BCB-R046 &                       8,20,21,23 \\
--     & --                 & --       & --          &
 05 25 30.6 & $-$70 09 13 &
 {\bf SHV0526001$-$701142}           &                            20,21 \\
--     & --                 & --       & --          &
 05 26 17.4 & $-$69 08 07 &
 {\bf WBP14}                         &                               15 \\
--     & --                 & --       & --          &
 05 35 11.4 & $-$70 22 46 &
 {\bf SHV0535442$-$702433}           &                            20,21 \\
\hline
\multicolumn{8}{c}{{\em non-IRAS sources classified as M or S stars}} \\
\hline
--     & --                 & --       &  {\bf 2446} &
 05 20 01.5 & $-$67 34 43 &
 WOH G274, GRV0520$-$6737            &                            11,18 \\
--     & --                 & --       & --          &
 05 21 33.1 & $-$70 09 56 &
 {\bf SHV0522023$-$701242}           &                            20,21 \\
--     & --                 & --       & --          &
 05 21 40.5 & $-$70 22 31 &
 {\bf SHV0522118$-$702517}           &                            20,21 \\
--     & --                 & --       & --          &
 05 24 31.3 & $-$69 43 25 &
 {\bf SHV0524565$-$694559}           &                            20,21 \\
--     & --                 & --       & --          &
 05 30 00.3 & $-$70 20 06 &
 {\bf SHV0530323$-$702216}           &                         20,21,23 \\
--     & --                 & --       & {\bf 12070} &
 05 52 27.8 & $-$69 14 12 &
 WOH SG515                           &                             9,11 \\
\hline
\multicolumn{8}{c}{{\em non-IRAS sources without spectral classification}} \\
\hline
--     & --                 & --       & --          &
 05 00 11.2 & $-$68 12 48 &
 {\bf SHV0500193$-$681706}           &                            20,21 \\
--     & --                 & --       & --          &
 05 00 13.5 & $-$68 24 56 &
 {\bf SHV0500233$-$682914}           &                            20,21 \\
--     & --                 & --       & --          &
 05 19 41.8 & $-$66 57 50 &
 {\bf GRV0519$-$6700}                &                               18 \\
\hline
\end{tabular}
\end{flushleft}
\normalsize
\end{table*}

The sources observed with ISO were selected from the lists presented in Paper
I, where all IRAS candidate AGB stars in the MCs are listed. We selected 30
infrared AGB stars or RSGs without optical counterparts from their Table 2.
These objects should represent the brightest, most obscured AGB stars. Four
objects from this table were excluded because of their red IRAS colours
($S_{25}/S_{12}\gsim2.5$): LI-LMC528, 861, 1137 and 1341. We also selected 8
sources from the optically known M and C stars with IRAS counterparts in Table
1 of Paper I. These include well known Harvard Variables as well as the
optically thick source IRAS04553$-$6825 (LI-LMC181, WOH G64; Elias et al.\
1986; Wood et al.\ 1986). Two unidentified IRAS sources from Table 4 of Paper
I have been included in the present sample. LI-LMC203 is near an M1.5 star
(HV12501), but there is also an A3 Iab supergiant (Sk$-69$-39a) close to the
IRAS position. For LI-LMC1795 we found a bright R-band counterpart (Paper II).
Finally one source from Table 7 of Paper I was included (LI-LMC1130). Although
listed in Paper I as a foreground star, it was included here in an attempt to
establish whether this is true. For these last three stars the higher spatial
resolution ISO observations at 12 $\mu$m allow a better identification of the
source with one of the possible counterparts found near the IRAS position. The
41 IRAS sources included in this study are listed in Table 1, with the most
common names for these objects, their coordinates (J2000) and some references.
The coordinates for the pointings of the ISO observations were taken from the
SIMBAD astronomical database in 1994.

The selection of IRAS detected AGB stars gives a sample that is severely
biased towards very luminous stars (including supergiants). We therefore also
included 16 non-IRAS stars. These were mostly taken from Wood et al.\ (1983,
1985), Reid et al.\ (1990) and Hughes (1989). Seven of these objects are
classified as C stars from optical spectra or near-IR colours. Six objects are
classified as M or S stars and for three objects no classification is
available. This group of non-IRAS sources also includes the RCB-like variable
HV2379 (Bessell \& Wood 1983). These sources are listed in Table 2.

\section{IRAS data}

%
% TABLE 3
%
\begin{table}
\caption[]{Revised IRAS 12, 25, 60 and 100 $\mu$m photometry (in Jy),
accompanied by a colon if questionable.}
\begin{tabular}{lllll}
\hline\hline
Star                           &
$F_{12}$                       &
$F_{25}$                       &
$F_{60}$                       &
$F_{100}$                      \\
\hline
GRV0519$-$6700                 &
 \llap{$<$}0.06                &
 \llap{$<$}0.02                &
 \llap{$<$}0.1                 &
                               \\
HV12070                        &
           0.06                &
           0.03                &
           0.1\rlap{:}         &
                               \\
HV12501                        &
           0.23                &
           0.06                &
 \llap{$<$}1.0                 &
                               \\
HV2360                         &
           0.38                &
           0.35                &
           0.4\rlap{:}         &
                               \\
HV2379                         &
           0.05                &
           0.02                &
 \llap{$<$}0.8                 &
                               \\
HV2446                         &
           0.05                &
           0.02                &
 \llap{$<$}0.2                 &
                               \\
HV5870                         &
           0.30                &
           0.17                &
 \llap{$<$}5.0                 &
                               \\
HV888                          &
           0.58                &
           0.29                &
 \llap{$<$}4.0                 &
                               \\
HV916                          &
           0.44                &
           0.23                &
 \llap{$<$}2.0                 &
                               \\
HV996                          &
           0.71                &
           0.53                &
 \llap{$<$}0.5                 &
                               \\
IRAS04286$-$6937               &
           0.28                &
           0.20                &
 \llap{$<$}0.1                 &
                               \\
IRAS04374$-$6831               &
           0.24                &
           0.12                &
           0.1\rlap{:}         &
                               \\
IRAS04407$-$7000               &
           0.76                &
           0.76                &
           0.1                 &
                               \\
IRAS04496$-$6958               &
           0.31                &
           0.19                &
           0.1                 &
                               \\
IRAS04498$-$6842               &
           1.33                &
           0.89                &
 \llap{$<$}0.2                 &
                               \\
IRAS04509$-$6922               &
           0.89                &
           0.86                &
 \llap{$<$}2.0                 &
                               \\
IRAS04516$-$6902               &
           0.86                &
           0.55                &
           0.4\rlap{:}         &
                               \\
IRAS04530$-$6916               &
           2.07                &
           5.09                &
   \llap{2}2.0                 &
   \llap{2}8.0                 \\
IRAS04539$-$6821               &
           0.22                &
           0.12                &
           0.1\rlap{:}         &
                               \\
IRAS04545$-$7000               &
           0.46                &
           0.83                &
 \llap{$<$}0.5                 &
                               \\
IRAS04557$-$6753               &
           0.24                &
           0.14                &
 \llap{$<$}0.3                 &
                               \\
IRAS05003$-$6712               &
           0.43                &
           0.33                &
           0.1\rlap{:}         &
                               \\
IRAS05009$-$6616               &
           0.28                &
           0.14                &
 \llap{$<$}0.4                 &
                               \\
IRAS05112$-$6755               &
           0.46                &
           0.33                &
           0.7\rlap{:}         &
                               \\
IRAS05113$-$6739               &
           0.25                &
           0.14                &
           0.1\rlap{:}         &
                               \\
IRAS05128$-$6455               &
           0.23                &
           0.24                &
           0.1                 &
                               \\
IRAS05190$-$6748               &
           0.39                &
           0.25                &
           0.1\rlap{:}         &
                               \\
IRAS05289$-$6617               &
           0.16                &
           0.39                &
           0.3                 &
                               \\
IRAS05294$-$7104               &
           0.69                &
           0.56                &
 \llap{$<$}3.0                 &
                               \\
IRAS05295$-$7121               &
           0.23                &
           0.08                &
 \llap{$<$}0.3                 &
                               \\
IRAS05298$-$6957               &
           0.85                &
           1.38                &
 \llap{$<$}3.0                 &
                               \\
IRAS05300$-$6651               &
           0.28                &
           0.17                &
           0.1\rlap{:}         &
                               \\
IRAS05329$-$6708               &
           0.74                &
           1.23                &
           0.2\rlap{:}         &
                               \\
IRAS05348$-$7024               &
           0.58                &
           0.16                &
 \llap{$<$}1.0                 &
                               \\
IRAS05360$-$6648               &
           0.21                &
           0.09                &
           0.3\rlap{:}         &
                               \\
IRAS05402$-$6956               &
           0.71                &
           1.02                &
 \llap{$<$}2.0                 &
                               \\
IRAS05506$-$7053               &
           0.28                &
           0.16                &
 \llap{$<$}0.2                 &
                               \\
IRAS05558$-$7000               &
           0.85                &
           0.80                &
           0.2\rlap{:}         &
                               \\
IRAS05568$-$6753               &
           0.35                &
           0.43                &
           0.2                 &
                               \\
SHV0454030$-$675031            &
 \llap{$<$}0.03                &
 \llap{$<$}0.03                &
 \llap{$<$}0.2                 &
                               \\
SHV0500193$-$681706            &
           0.11\rlap{:}        &
           0.07                &
 \llap{$<$}0.3                 &
                               \\
SHV0500233$-$682914            &
           0.10\rlap{:}        &
           0.03                &
 \llap{$<$}1.5                 &
                               \\
SHV0502469$-$692418            &
           0.02\rlap{:}        &
 \llap{$<$}0.03                &
 \llap{$<$}0.1                 &
                               \\
SHV0521050$-$690415            &
           0.06\rlap{:}        &
           0.02\rlap{:}        &
 \llap{$<$}0.7                 &
                               \\
SHV0522023$-$701242            &
 \llap{$<$}0.10                &
 \llap{$<$}0.04                &
           0.4\rlap{:}         &
                               \\
SHV0522118$-$702517            &
           0.06\rlap{:}        &
           0.05                &
 \llap{$<$}2.2                 &
                               \\
SHV0524565$-$694559            &
 \llap{$<$}0.14                &
 \llap{$<$}0.07                &
 \llap{$<$}1.0                 &
                               \\
SHV0526001$-$701142            &
           0.07\rlap{:}        &
           0.01\rlap{:}        &
           0.1\rlap{:}         &
                               \\
SHV0530323$-$702216            &
 \llap{$<$}0.04                &
 \llap{$<$}0.04                &
           0.4                 &
                               \\
SHV0535442$-$702433            &
           0.01\rlap{:}        &
           0.07\rlap{:}        &
 \llap{$<$}1.0                 &
                               \\
SP77 30$-$6                    &
           0.26                &
           0.13                &
           0.1\rlap{:}         &
                               \\
TRM45                          &
           0.07                &
           0.07                &
 \llap{$<$}2.0                 &
                               \\
TRM72                          &
           0.22                &
           0.06                &
 \llap{$<$}0.3                 &
                               \\
TRM88                          &
           0.17                &
           0.04                &
 \llap{$<$}0.7                 &
                               \\
WBP14                          &
           0.01\rlap{:}        &
 \llap{$<$}0.03                &
 \llap{$<$}4.0                 &
                               \\
WOH G64                        &
           8.45                &
   \llap{1}3.53                &
           2.2                 &
                               \\
WOH SG374                      &
           0.37                &
           0.38                &
           0.2\rlap{:}         &
                               \\
\hline
\end{tabular}
\end{table}

Here the IRAS data are discussed, for later comparison with the ISO
photometry. Data at 12, 25, 60 and 100 $\mu$m was retrieved from the IRAS data
base server in Groningen\footnote{The IRAS data base server of the Space
Research Organisation of the Netherlands (SRON) and the Dutch Expertise Centre
for Astronomical Data Processing is funded by the Netherlands Organisation for
Scientific Research (NWO). The IRAS data base server project was also partly
funded through the Air Force Office of Scientific Research, grants
AFOSR 86-0140 and AFOSR 89-0320.} (Assendorp et al.\ 1995). The Groningen
Gipsy data analysis software was used to measure the flux density from a trace
through the position of the star (Gipsy command SCANAID). For the 60 and 100
$\mu$m data, $2\times2$ square degree maps were created with $0.5^\prime$
pixels to find point sources coincident with the positions of the stars. The
12 and 25 $\mu$m flux densities have a 1-$\sigma$ error of a few per cent,
with a minimum error of $\sim0.01$ Jy. The 60 and 100 $\mu$m flux densities
are much less certain, and it is also more difficult to assess reliable error
estimates: 10\% would be a typical error. The faintest 60 $\mu$m sources that
IRAS detected were assigned $F_{60}=0.1$ Jy. Only one source was well detected
at 100 $\mu$m. The flux densities are listed in Table 3. When it is not
certain that the measured flux density is physically related to the star of
interest it is marked with a colon.

All of our sources that are in the IRAS-PSC, plus HV5870 (=LI-LMC1145) and
TRM72 (=LI-LMC578) that are in Schwering \& Israel (1990), were recovered with
good flux density determinations at 12 $\mu$m. Reliable 12 $\mu$m flux
densities could also be determined for IRAS05128$-$6455 and 05289$-$6617,
below their upper limits as listed in the PSC. Neither in the PSC, nor in
Schwering \& Israel (1990), are HV12070, HV2379, HV2446, TRM45, and TRM88
secure detections. Detection is not certain for WBP14 and the SHV sources for
which flux density estimates are listed. The 12 $\mu$m flux densities of the
GRV source and four SHV sources are upper limits. IRAS05506$-$7053 looks
extended or multiple.

At 25 $\mu$m detections seem a little more reliable than at 12 $\mu$m, at a
given flux density. Rather surprisingly, the detection limit at 25 $\mu$m is
at least as faint as at 12 $\mu$m; sources with $F_{25}\sim0.02$ Jy could be
found (see also Reid et al.\ 1990). This is, however, only possible because
the positions of the stars are known. For SP77 30$-$6 and all eight (other)
IRAS sources the PSC lists only upper limits of $F_{25}<0.25$ Jy.
SHV0502469$-$692418 and WBP14 were the only sources that were (tentatively)
detected at 12 $\mu$m but not at 25 $\mu$m. Their flux densities are probably
below the limit of detection, $F_{25}<0.01$ Jy, if their colours are rather
blue.

At 60 $\mu$m, IRAS05298$-$6957 is a bright, small but extended source about
$10^\prime$ in diameter, with $F_{60}\sim2$ Jy. No point source could be
distinguished on top of this emission, that is probably associated with the
small cluster of which IRAS05298$-$6957 is a member (Paper IV). Flux densities
are listed for two dozen sources, but it is not sure how many among these are
real detections and how many are spurious. The only 60 $\mu$m detections in
the PSC are IRAS04516$-$6902 ($0.80\pm0.19$ Jy), 04530$-$6916 ($20.51\pm1.85$
Jy) and 05112$-$6755 ($0.91\pm0.11$ Jy), all consistent with our estimates.
More stringent upper limits are put on the 60 $\mu$m flux densities of the
other sources.

At 100 $\mu$m sources may be detected as faint as a few Jy. The only
detection, however, is the brightest far-IR source in our sample,
IRAS04530$-$6916, which we measured at $F_{100}=28$ Jy. This is consistent
with the PSC upper limit of 46.17 Jy.

The new flux density estimates can be compared with the literature values from
the PSC or Schwering \& Israel (1990) (Fig.\ 1). On average, the new flux
densities are only a few per cent fainter than the values from the literature.
Flux densities $F_{25}\lsim0.2$ Jy may have been over-estimated in the past.
HV12501 with $F_{\rm rev}$/$F_{\rm lit}=0.56$ and 0.32 at 12 and 25 $\mu$m,
respectively, and IRAS05506$-$7053 with $F_{\rm rev}$/$F_{\rm lit}=0.67$ and
0.42 at 12 and 25 $\mu$m, respectively, are the most extreme examples of this.
Schwering \& Israel (1990) over-estimated the 25 $\mu$m flux density of TRM72
($F_{\rm rev}$/$F_{\rm lit}=0.55$), but under-estimated the 12 $\mu$m flux
density of HV5870 ($F_{\rm rev}$/$F_{\rm lit}=2.00$). The other two flux
densities which are obviously under-estimated are for SP77 30$-$6 at 12 $\mu$m
($F_{\rm rev}$/$F_{\rm lit}=1.53$) and IRAS04286$-$6937 at 25 $\mu$m ($F_{\rm
rev}$/$F_{\rm lit}=1.67$).

%
% FIGURE 1
%
\begin{figure}[tb]
\centerline{\psfig{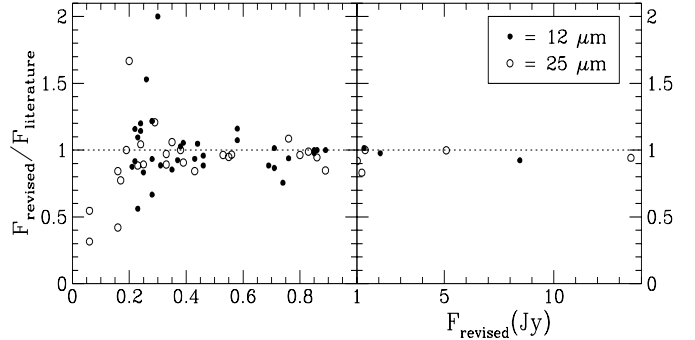}}
\caption[]{A comparison of the estimates of IRAS flux densities given here and
values from the IRAS Point Source Catalogue and Schwering \& Israel (1990), at
12 (solid symbols) and 25 $\mu$m (open symbols).}
\end{figure}

IRAS counterparts not listed in IRAS-based catalogues may still be found in
the original IRAS data. This is because manual extraction and measuring of the
data is a more sophisticated technique than the automatic techniques that
created the existing catalogues. In particular, manual flux-density
determination enables the background flux levels to be estimated and
subtracted better, yielding more reliable photometry. The only sources from
our ISO sample that could not be detected in the IRAS data at either 12, 25 or
60 $\mu$m are GRV0519$-$6700, SHV0454030$-$675031 and SHV0524565$-$694559.

\section{ISO observations}

The programme stars were observed with ISO at 12, 25 and 60 $\mu$m (chopped
measurements) and with PHOT-S as part of a Guaranteed Time programme under
proposals NTMCAGB1 and NTMCAGB2, and at 60 $\mu$m with mapping mode and
CAM-CVF as part of an Open Time programme under proposal LMCSPECT.

The 12 $\mu$m photometry was mostly obtained using the ISOCAM instrument
(Cesarsky et al.\ 1996) in staring mode with a $3^{\prime\prime}$ pixel field
of view in beam LW-s and using the LW10 filter ($\sim$ IRAS 12 $\mu$m). We did
25 exposures of each 2.1 s duration, after a number of read-outs to stabilise
the detector (ranging from 10 to 34 frames depending on the expected source
flux density). The gain was 2 in most cases, but 1 in the case of sources that
were expected to be relatively bright: HV12501 and 996, IRAS04496$-$6958,
04545$-$7000, 05003$-$6712, 05112$-$6755, 05348$-$7024 and 05568$-$6753, and
WOH SG374. For most stars this resulted in a clear detection with S/N ratios
of 10 to 100. In total 44 sources were observed with ISOCAM at 12 $\mu$m.

For sources that were expected to be stronger than 0.5 Jy and which would
therefore saturate the ISOCAM detectors with the LW10 filter, the 12 $\mu$m
photometry was obtained with the ISOPHOT instrument (Lemke et al.\ 1996) using
the 11.5 filter ($\sim$ IRAS 12 $\mu$m). These observations were done using
triangular chopping with a chopper throw of $90^{\prime\prime}$. The aperture
used for the observations was $52^{\prime\prime}$ in diameter. Integration
times were 32 s on-source (and the same off-source), except for
IRAS05294$-$7104 that we integrated 64 s. A total of 13 sources were observed
in this mode. For 53 sources we obtained PHOT-P photometry at 25 $\mu$m using
the 25 filter ($\sim$ IRAS 25 $\mu$m), triangular chopping with a chopper
throw of $90^{\prime\prime}$, and an aperture of $52^{\prime\prime}$.
Integration times ranged from 32 to 256 s, depending on the expected flux
density. In our Guaranteed Time programme we finally observed 40 objects with
ISOPHOT at 60 $\mu$m using the PHOT-C100 camera and filter 60 ($\sim$ IRAS 60
$\mu$m) and triangular chopping with a $150^{\prime\prime}$ chopper throw.
Integration times ranged from 32 to 128 s, depending on the expected flux
density. Unfortunately due to the reduced in-orbit sensitivity of the
instrument and the problems with the calibration of the chopped measurements
(especially for PHOT-C), we discovered after most observations had already
been carried out that this was not the best observing strategy for the 60
$\mu$m photometry. Therefore, 7 objects were observed again in the Open Time
using PHOT-C100 and filter 60 in raster mapping mode, with $3\times3$ rasters
and $45^{\prime\prime}$ raster steps in X and Y directions (spacecraft
coordinates). The integration time per raster point was 128 seconds.

In order to establish the carbon- or oxygen-rich nature of some of the
programme stars we also obtained IR spectra for a number of them. In the
Guaranteed Time 15 objects were observed using PHOT-S in staring mode, with
integration times of 256 or 512 s (1024 s for HV2379) depending on the
expected flux densities. The advantage of this instrument is that its spectral
coverage is rather large (2 to 12 $\mu$m) at a reasonable resolution
($\sim90$). The sensitivity of the PHOT-S instrument, however, limits the
detectability to sources with 12 $\mu$m flux densities above $\sim0.3$ Jy.
Furthermore, using staring observations the background cannot easily be
determined. In this spectral region the diffuse emission is dominated by the
zodiacal emission, which, according to IRAS measurements, amounts to about
$\sim0.1$ Jy in the PHOT-S aperture at 10 $\mu$m.

Considering this, we decided to obtain CAM-CVF spectra for 12 objects with a
pixel field-of-view of $6^{\prime\prime}$ in beam LW-l. We did 25 exposures of
2.1 s each at gain 2, after 50 read-outs to stabilise the detector. The
unprecedented sensitivity of the ISOCAM instrument allows the observer to
obtain spectra even for sources as faint as 100 mJy at 12 $\mu$m. Because of
the long duration of a CVF observation, the spectral coverage chosen was only
7 to 9.2 $\mu$m (with step 4) in LW-CVF1 and 9 to 14.1 $\mu$m (with step 2) in
LW-CVF2, at a spectral resolution of $\sim40$. A big advantage of the CAM-CVF
is that the spectra are obtained using an imaging technique. Therefore, a
background spectrum was obtained simultaneously. These background spectra can
be used to correct the PHOT-S spectra. We also obtained observations of 3
objects for which PHOT-S spectra had already been taken, in order to
cross-check the results from the different instruments.

\subsection{Near-IR photometry}

Near-IR photometry was determined for each star at the time of the ISO
observation, by interpolating near-IR lightcurves from our monitoring campaign
at the South African Astronomical Observatory (SAAO) at Sutherland, South
Africa (Whitelock et al., in preparation). Nearly always the lightcurve was
sampled close in time to the ISO observation, but occasionally some
extrapolation was necessary. The quoted uncertainties include an estimate, for
each star, of the error introduced by the inter/extrapolation. For TRM45 and
for the H-band magnitude of IRAS05360$-$6648 we have made use of the near-IR
lightcurves and photometry presented by Wood (1998), after transformation to
the SAAO system using Carter (1990). The near-IR photometry is listed in Table
4, along with the ISO photometry and the Julian Dates of the ISO spectroscopy.

No near-IR counterparts could be identified with IRAS05568$-$6753 and
05289$-$6617. Two stars with near-IR colours much like those of unobscured
M-type stars were monitored in the near-IR, but they show no variability.

\section{ISO results and comparison with IRAS photometry}

The data were reduced using the PHOT and CAM Interactive Analysis software
packages: PIA (Gabriel et al.\ 1997) version V7.1.2(e) and CIA (Ott et al.\
1996) version V3.0, respectively. For a general description of the data and
reduction methods see the ISOPHOT Data Users Manual (Laureijs et al.\ 1998),
and the ISOCAM Observer's Manual (1994) and ISOCAM Data Users Manual
(Siebenmorgen et al.\ 1998). Details of the steps undertaken in reducing
so-called Edited Raw Data (ERD) products to the finally derived flux densities
and spectra can be found in Appendices A (photometry) and B (spectroscopy).
The resulting ISO photometry is listed in Table 4, and the ISO spectra are
presented in Figs.\ 4 \& 5.

The flux densities at 12 and 25 $\mu$m for the stars that were detected both
by IRAS and ISO (Tables 3 \& 4) are compared in Fig.\ 2. A bright regime where
ISO and IRAS are consistent can be distinguished from a faint regime where ISO
flux densities are systematically lower than IRAS flux densities. CAM is
consistent with IRAS down to fainter levels ($\sim0.2$ Jy) than PHOT
($\sim0.6$ Jy). PHOT seems to under-estimate flux densities at levels between
0.2 and 0.6 Jy by a factor $\sim$two. Below 0.2 Jy, both CAM and PHOT yield
flux densities $\sim0.6\times$IRAS.

%
% FIGURE 2
%
\begin{figure}[tb]
\centerline{\psfig{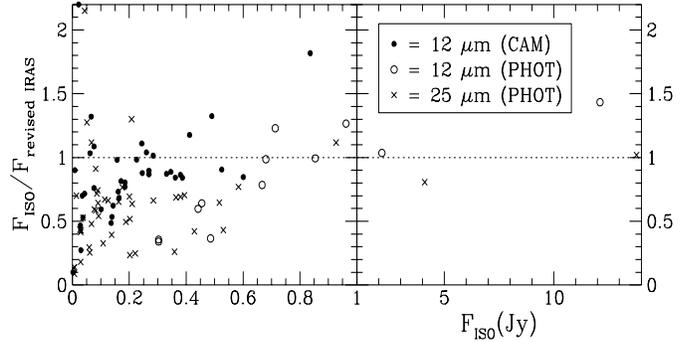}}
\caption[]{Comparison between ISO and our revised IRAS flux densities.}
\end{figure}

Flux density under-estimation may be caused by the difficulty of the detectors
to respond to low signals. CAM 12 $\mu$ flux densities below $\sim0.2$ Jy may
have been under-estimated as the $\kappa$-$\sigma$ method does not adequately
correct for non-stabilised signals if stabilisation is not reached well within
less than half the integration time. The 12 and 25 $\mu$m PHOT measurements,
although employing different detector materials, show exactly the same trend.
The 12 (either CAM or PHOT) and 25 $\mu$m observations were performed in the
same orbit. Default responsivities for the 12 and 25 $\mu$m PHOT measurements
were thought to be (much) lower in 1996 than in the current calibrations.
Adopting those early values, the ISOPHOT photometry would be consistent with
the IRAS photometry to a high degree. Subsequent revisions of the default
responsivities have lead to higher values, approaching the values we obtain
from the (chopped) FCS measurements. The ratio of ISO and IRAS flux densities
at 25 $\mu$m is $0.60\pm0.06$ for the 24 sources with ISO measurements before
orbit 190, and $0.68\pm0.09$ for the 21 sources measured after orbit 190.
These ratios are very similar, despite the large differences in median IRAS
flux density between these two samples: 0.41 and 0.14 Jy, respectively.

However, the discrepancy between the ISO and IRAS data may not be as great as
it appears if we take plausible selection effects into account. The stars in
our sample were largely selected on the basis of their IRAS flux density, but
many of them were only just detected by IRAS. It is therefore likely that they
were near the maximum of their pulsation cycles at the time of the IRAS
observation. In contrast, they will have been at random phases when the ISO
observations were made. This will lead to a systematic difference between
the IRAS and ISO flux densities for faint sources. A similar effect may
explain the discrepancy between the PHOT and CAM behaviour for sources with
flux densities in the range 0.2 to 0.6 Jy, as the brighter sources were
selected for measurement with PHOT and the fainter sources with CAM.
Ground-based 10 $\mu$m (N-band) magnitudes of a subset of our ISO targets were
on average $\sim30$\% fainter than measured by IRAS at 12 $\mu$m (Paper IV).
Although we explained this in terms of differences between the N-band and IRAS
12 $\mu$m filters, it may actually reflect the same discrepancy seen between
the ISO and IRAS photometry. Variability cannot be the complete explanation,
though: for instance, the sources IRAS04407$-$7000, 4516$-$6902 and
05003$-$6712 were all near the maxima in their K- and L-band lightcurves at
the time of the ISO photometry, yet their PHOT 25 $\mu$m flux densities of
0.584, 0.380, and 0.210 Jy, respectively, are still fainter than the IRAS flux
densities by factors of 0.77, 0.69, and 0.64, respectively. Interestingly,
Reid et al.\ (1990, their Figure 5) show that for $F_{12,25}\lsim0.3$ Jy both
the PSC and Schwering \& Israel (1990) over-estimate flux densities for point
sources in the LMC by typically 20 to 50\%. They attribute this to source
confusion resulting from the large beam width and crowdedness in typical LMC
fields.

%
% TABLE 4a
%
\begin{landscape}
\begin{table}
\caption[]{ISO 12, 25 and 60 $\mu$m photometry (in Jy). The near-IR magnitudes
for the ISO-epochs is deduced from light curves obtained at SAAO
($JD-2,450,000={\rm orbit}+38$). Values in parentheses are 1-$\sigma$ errors.
The last column indicates when a PHOT-S or CAM-CVF spectrum was taken.}
\begin{tabular}{lllllllllllll}
\hline\hline
Star                           &
 $JD$                          &
 $J [mag]$                     &
 $H [mag]$                     &
 $K [mag]$                     &
 $L [mag]$                     &
 $F_{12}$(CAM)                 &
 $F_{12}$(PHOT)                &
 $F_{25}$(PHOT)                &
 $F_{60}$(chop)                &
 $F_{60}$(map)                 &
 Spectrum                      \\
\hline
GRV0519$-$6700                 &
 318                           &
 \llap{1}2.63(0.03)            &
 \llap{1}1.36(0.02)            &
 \llap{1}0.67(0.02)            &
                               &
           0.004(0.001)        &
                               &
                               &
                               &
                               &
                               \\
HV12070                        &
 274                           &
 \llap{1}0.03(0.02)            &
         8.98(0.02)            &
         8.68(0.02)            &
         8.24(0.03)            &
           0.043(0.004)        &
                               &
 \llap{$-$}0.023(0.003)        &
                               &
                               &
                               \\
                               &
 787                           &
 \llap{1}0.30(0.10)            &
         9.20(0.10)            &
         8.80(0.10)            &
         8.30(0.10)            &
                               &
                               &
                               &
                               &
                               &
 CAM-CVF                       \\
HV12501                        &
 754                           &
         8.80(0.05)            &
         8.00(0.05)            &
         7.75(0.05)            &
         7.30(0.05)            &
           0.185(0.005)        &
                               &
           0.067(0.018)        &
           0.260(0.108)        &
                               &
                               \\
HV2360                         &
 754                           &
         9.00(0.05)            &
         8.15(0.05)            &
         7.75(0.05)            &
         7.20(0.05)            &
           0.331(0.003)        &
                               &
           0.138(0.019)        &
           0.207(0.042)        &
                               &
                               \\
HV2379                         &
 288                           &
 \llap{1}2.68(0.03)            &
 \llap{1}2.05(0.04)            &
 \llap{1}1.40(0.04)            &
         9.70(0.10)            &
           0.038(0.002)        &
                               &
                               &
 \llap{$-$}0.027(0.048)        &
                               &
 PHOT-S                        \\
                               &
 626                           &
                               &
 \llap{1}3.20(0.40)            &
 \llap{1}1.90(0.20)            &
                               &
           0.035(0.001)        &
                               &
                               &
                               &
                               &
                               \\
                               &
 729                           &
                               &
 \llap{1}3.20(0.40)            &
 \llap{1}1.90(0.20)            &
                               &
           0.032(0.001)        &
                               &
           0.014(0.005)        &
                               &
                               &
 CAM-CVF                       \\
HV2446                         &
 288                           &
 \llap{1}0.25(0.02)            &
         9.23(0.02)            &
         8.80(0.02)            &
         8.35(0.04)            &
           0.066(0.001)        &
                               &
           0.043(0.005)        &
           0.109(0.056)        &
                               &
                               \\
                               &
 622                           &
 \llap{1}0.60(0.10)            &
         9.60(0.10)            &
         9.10(0.10)            &
         8.35(0.05)            &
                               &
                               &
                               &
                               &
           0.361(0.135)        &
                               \\
                               &
 754                           &
 \llap{1}0.10(0.02)            &
         9.04(0.02)            &
         8.67(0.02)            &
         8.06(0.02)            &
                               &
                               &
                               &
                               &
                               &
 CAM-CVF                       \\
HV5870                         &
 754                           &
         9.34(0.02)            &
         8.40(0.02)            &
         8.10(0.02)            &
         7.60(0.04)            &
           0.269(0.002)        &
                               &
           0.092(0.008)        &
           0.090(0.355)        &
                               &
                               \\
HV888                          &
 195                           &
         8.10(0.01)            &
         7.20(0.01)            &
         6.89(0.01)            &
         6.46(0.02)            &
                               &
           0.713(0.037)        &
           0.201(0.012)        &
 \llap{$-$}0.237(0.626)        &
                               &
 PHOT-S                        \\
HV916                          &
 209                           &
         8.55(0.03)            &
         7.60(0.03)            &
         7.25(0.02)            &
         6.80(0.02)            &
           0.380(0.004)        &
                               &
           0.176(0.015)        &
           0.065(0.171)        &
                               &
                               \\
HV996                          &
 163                           &
         8.93(0.01)            &
         8.01(0.01)            &
         7.58(0.01)            &
         6.83(0.01)            &
           0.601(0.006)        &
                               &
           0.364(0.026)        &
           0.326(0.074)        &
                               &
 PHOT-S                        \\
IRAS04286$-$6937               &
 217                           &
                               &
 \llap{1}3.00(0.05)            &
 \llap{1}1.25(0.05)            &
         9.10(0.05)            &
           0.136(0.002)        &
                               &
           0.059(0.013)        &
 \llap{$-$}0.101(0.130)        &
                               &
                               \\
IRAS04374$-$6831               &
 195                           &
                               &
 \llap{1}3.60(0.05)            &
 \llap{1}1.40(0.04)            &
         8.80(0.03)            &
           0.185(0.003)        &
                               &
           0.086(0.027)        &
           0.228(0.136)        &
                               &
 PHOT-S                        \\
                               &
 701                           &
                               &
 \llap{1}3.60(0.10)            &
 \llap{1}1.40(0.10)            &
         8.90(0.10)            &
                               &
                               &
                               &
                               &
           0.309(0.069)        &
                               \\
IRAS04407$-$7000               &
 217                           &
 \llap{1}0.18(0.03)            &
         8.92(0.03)            &
         8.18(0.03)            &
         7.30(0.05)            &
                               &
           0.962(0.024)        &
           0.584(0.031)        &
           0.125(0.088)        &
                               &
                               \\
                               &
 605                           &
 \llap{1}1.70(0.05)            &
 \llap{1}0.20(0.05)            &
         9.20(0.05)            &
         8.10(0.05)            &
                               &
                               &
                               &
                               &
           0.473(0.132)        &
                               \\
IRAS04496$-$6958               &
 195                           &
 \llap{1}3.00(0.05)            &
 \llap{1}0.90(0.05)            &
         9.50(0.04)            &
         7.70(0.04)            &
           0.269(0.002)        &
                               &
           0.126(0.010)        &
           0.252(0.154)        &
                               &
 PHOT-S                        \\
                               &
 605                           &
 \llap{1}2.40(0.05)            &
 \llap{1}0.40(0.05)            &
         8.95(0.05)            &
         7.60(0.05)            &
                               &
                               &
                               &
                               &
                               &
 CAM-CVF                       \\
IRAS04498$-$6842               &
 217                           &
 \llap{1}0.95(0.05)            &
         9.65(0.05)            &
         8.70(0.02)            &
         7.70(0.02)            &
                               &
           0.486(0.013)        &
           0.221(0.028)        &
           0.292(0.112)        &
                               &
                               \\
IRAS04509$-$6922               &
 217                           &
 \llap{1}1.50(0.05)            &
 \llap{1}0.00(0.04)            &
         9.15(0.04)            &
         8.10(0.04)            &
                               &
           0.303(0.011)        &
           0.202(0.028)        &
 \llap{$-$}0.021(0.231)        &
                               &
                               \\
IRAS04516$-$6902               &
 202                           &
 \llap{1}0.50(0.03)            &
         8.96(0.03)            &
         8.24(0.03)            &
         7.25(0.03)            &
                               &
           0.854(0.026)        &
           0.380(0.051)        &
           0.349(0.135)        &
                               &
                               \\
IRAS04530$-$6916               &
 202                           &
 \llap{1}3.65(0.05)            &
 \llap{1}1.60(0.04)            &
         9.73(0.02)            &
         7.60(0.03)            &
                               &
           2.144(0.079)        &
           4.105(0.054)        &
   \llap{3}7.656(2.639)        &
                               &
                               \\
IRAS04539$-$6821               &
 229                           &
                               &
 \llap{1}4.30(0.05)            &
 \llap{1}1.80(0.04)            &
         8.80(0.10)            &
           0.244(0.002)        &
                               &
           0.089(0.013)        &
           0.106(0.053)        &
                               &
                               \\
IRAS04545$-$7000               &
 195                           &
                               &
 \llap{1}1.75(0.03)            &
         9.40(0.02)            &
         7.15(0.03)            &
           0.836(0.008)        &
                               &
           0.927(0.027)        &
           0.641(0.192)        &
                               &
 PHOT-S                        \\
                               &
 605                           &
                               &
 \llap{1}2.80(0.10)            &
 \llap{1}0.30(0.10)            &
         8.10(0.10)            &
                               &
                               &
                               &
                               &
           0.286(0.206)        &
                               \\
IRAS04557$-$6753               &
 229                           &
                               &
 \llap{1}4.60(0.05)            &
 \llap{1}2.45(0.04)            &
         9.60(0.05)            &
           0.164(0.002)        &
                               &
           0.090(0.027)        &
 \llap{$-$}0.072(0.137)        &
                               &
                               \\
IRAS05003$-$6712               &
 229                           &
 \llap{1}2.30(0.05)            &
 \llap{1}0.65(0.05)            &
         9.45(0.05)            &
         8.15(0.05)            &
           0.362(0.003)        &
                               &
           0.210(0.007)        &
           0.173(0.079)        &
                               &
 PHOT-S                        \\
IRAS05009$-$6616               &
 229                           &
 \llap{1}4.70(0.10)            &
 \llap{1}2.65(0.05)            &
 \llap{1}0.70(0.05)            &
         8.45(0.05)            &
           0.284(0.003)        &
                               &
           0.082(0.017)        &
 \llap{$-$}0.032(0.047)        &
                               &
                               \\
IRAS05112$-$6755               &
 288                           &
                               &
 \llap{1}4.70(0.10)            &
 \llap{1}2.00(0.05)            &
         8.80(0.05)            &
           0.414(0.004)        &
                               &
                               &
           0.078(0.133)        &
                               &
 PHOT-S                        \\
                               &
 629                           &
                               &
 \llap{1}4.70(0.10)            &
 \llap{1}1.90(0.05)            &
         8.65(0.05)            &
           0.360(0.003)        &
                               &
           0.108(0.011)        &
                               &
                               &
 PHOT-S                        \\
IRAS05113$-$6739               &
 288                           &
                               &
 \llap{1}4.35(0.05)            &
 \llap{1}2.12(0.03)            &
         9.06(0.02)            &
           0.316(0.003)        &
                               &
                               &
           0.398(0.115)        &
                               &
                               \\
                               &
 629                           &
                               &
 \llap{1}3.90(0.10)            &
 \llap{1}1.60(0.10)            &
         8.75(0.05)            &
           0.209(0.002)        &
                               &
                               &
                               &
                               &
                               \\
                               &
 729                           &
                               &
 \llap{1}3.90(0.05)            &
 \llap{1}1.60(0.05)            &
         8.75(0.05)            &
           0.254(0.002)        &
                               &
           0.067(0.013)        &
                               &
                               &
                               \\
IRAS05128$-$6455               &
 628                           &
 \llap{1}3.60(0.10)            &
 \llap{1}2.10(0.10)            &
 \llap{1}0.55(0.05)            &
         8.55(0.05)            &
           0.226(0.002)        &
                               &
           0.061(0.008)        &
                               &
                               &
 PHOT-S                        \\
IRAS05190$-$6748               &
 288                           &
                               &
                               &
 \llap{1}3.10(0.05)            &
         9.70(0.10)            &
           0.346(0.003)        &
                               &
           0.163(0.022)        &
           0.190(0.060)        &
                               &
 PHOT-S                        \\
IRAS05289$-$6617               &
 209                           &
                               &
                               &
                               &
                               &
           0.157(0.002)        &
                               &
           0.202(0.009)        &
           0.382(0.049)        &
                               &
                               \\
                               &
 668                           &
                               &
                               &
                               &
                               &
                               &
                               &
                               &
                               &
           0.281(0.069)        &
                               \\
                               &
 729                           &
                               &
                               &
                               &
                               &
                               &
                               &
                               &
                               &
                               &
 CAM-CVF                       \\
\hline
\end{tabular}
\end{table}
\end{landscape}

%
% TABLE 4b
%
\begin{landscape}
\begin{table}
\addtocounter{table}{-1}
\caption[]{(continued) The near-IR photometry of Wood (1998) is used for TRM45
and in part for IRAS05360$-$6648 (H-band), after transformation to the SAAO
system (Carter 1990).}
\begin{tabular}{llllllllllll}
\hline\hline
Star                           &
 $JD$                          &
 $J [mag]$                     &
 $H [mag]$                     &
 $K [mag]$                     &
 $L [mag]$                     &
 $F_{12}$(CAM)                 &
 $F_{12}$(PHOT)                &
 $F_{25}$(PHOT)                &
 $F_{60}$(chop)                &
 $F_{60}$(map)                 &
 Spectrum                      \\
\hline
IRAS05294$-$7104               &
 754                           &
 \llap{1}1.80(0.05)            &
 \llap{1}0.00(0.05)            &
         8.90(0.05)            &
         7.60(0.05)            &
                               &
           0.680(0.045)        &
           0.394(0.035)        &
           0.079(0.097)        &
                               &
                               \\
IRAS05295$-$7121               &
 209                           &
                               &
 \llap{1}3.85(0.05)            &
 \llap{1}1.75(0.05)            &
         9.35(0.05)            &
           0.143(0.001)        &
                               &
           0.007(0.011)        &
           0.233(0.074)        &
                               &
                               \\
IRAS05298$-$6957               &
 209                           &
                               &
                               &
 \llap{1}1.60(0.20)            &
         8.60(0.20)            &
                               &
           0.303(0.015)        &
           0.359(0.014)        &
           0.414(0.279)        &
                               &
 PHOT-S                        \\
                               &
 729                           &
 \llap{1}4.10(0.20)            &
 \llap{1}2.50(0.10)            &
 \llap{1}1.10(0.10)            &
         8.50(0.20)            &
                               &
                               &
                               &
                               &
                               &
 CAM-CVF                       \\
IRAS05300$-$6651               &
 209                           &
                               &
 \llap{1}4.70(0.20)            &
 \llap{1}2.20(0.10)            &
         9.10(0.10)            &
           0.246(0.003)        &
                               &
           0.114(0.017)        &
           0.068(0.049)        &
                               &
                               \\
IRAS05329$-$6708               &
 163                           &
 \llap{1}7.00(0.20)            &
 \llap{1}2.70(0.10)            &
 \llap{1}0.40(0.05)            &
         8.25(0.05)            &
                               &
           0.442(0.018)        &
           0.531(0.035)        &
           0.479(0.082)        &
                               &
 PHOT-S                        \\
IRAS05348$-$7024               &
 202                           &
                               &
 \llap{1}3.60(0.30)            &
 \llap{1}1.60(0.20)            &
         8.70(0.20)            &
           0.525(0.005)        &
                               &
           0.208(0.031)        &
           0.170(0.108)        &
                               &
                               \\
                               &
 732                           &
                               &
 \llap{1}3.80(0.30)            &
 \llap{1}2.70(0.20)            &
         9.30(0.20)            &
                               &
                               &
                               &
                               &
                               &
 CAM-CVF                       \\
IRAS05360$-$6648               &
 202                           &
                               &
 \llap{1}5.00(0.40)            &
 \llap{1}2.30(0.05)            &
         9.50(0.10)            &
           0.171(0.002)        &
                               &
           0.082(0.013)        &
           0.174(0.067)        &
                               &
                               \\
IRAS05402$-$6956               &
 217                           &
                               &
 \llap{1}3.50(0.10)            &
 \llap{1}0.60(0.05)            &
         8.00(0.04)            &
                               &
           0.455(0.020)        &
           0.429(0.022)        &
           0.944(0.288)        &
                               &
                               \\
                               &
 662                           &
 \llap{1}4.40(0.10)            &
 \llap{1}1.45(0.05)            &
         9.40(0.03)            &
         7.15(0.03)            &
                               &
                               &
                               &
                               &
 \llap{$-$}0.245(0.527)        &
                               \\
                               &
 782                           &
 \llap{1}4.50(0.10)            &
 \llap{1}1.50(0.05)            &
         9.50(0.05)            &
         7.25(0.05)            &
                               &
                               &
                               &
                               &
                               &
 CAM-CVF                       \\
IRAS05506$-$7053               &
 173                           &
                               &
 \llap{1}6.00(0.30)            &
 \llap{1}3.30(0.10)            &
 \llap{1}0.00(0.10)            &
                               &
                               &
                               &
           0.109(0.062)        &
                               &
                               \\
                               &
 228                           &
                               &
 \llap{1}6.40(0.30)            &
 \llap{1}3.60(0.10)            &
 \llap{1}0.30(0.10)            &
                               &
 \llap{$-$}0.035(0.017)        &
           0.029(0.015)        &
           0.167(0.073)        &
                               &
                               \\
IRAS05558$-$7000               &
 209                           &
 \llap{1}1.90(0.04)            &
 \llap{1}0.10(0.04)            &
         8.90(0.03)            &
         7.60(0.04)            &
                               &
           0.667(0.011)        &
           0.517(0.016)        &
           0.291(0.100)        &
                               &
 PHOT-S                        \\
                               &
 781                           &
 \llap{1}3.50(0.05)            &
 \llap{1}1.60(0.03)            &
 \llap{1}0.03(0.01)            &
         8.20(0.05)            &
                               &
                               &
                               &
                               &
                               &
 CAM-CVF                       \\
IRAS05568$-$6753               &
 186                           &
                               &
                               &
                               &
                               &
           0.412(0.004)        &
                               &
           0.285(0.008)        &
           0.383(0.073)        &
                               &
 PHOT-S                        \\
SHV0454030$-$675031            &
 318                           &
 \llap{1}4.20(0.10)            &
 \llap{1}2.67(0.04)            &
 \llap{1}1.85(0.03)            &
                               &
           0.006(0.002)        &
                               &
                               &
                               &
                               &
                               \\
SHV0500193$-$681706            &
 229                           &
 \llap{1}4.00(0.20)            &
 \llap{1}2.00(0.10)            &
 \llap{1}0.70(0.10)            &
         9.50(0.10)            &
           0.030(0.005)        &
                               &
           0.030(0.016)        &
           0.098(0.026)        &
                               &
                               \\
                               &
 754                           &
 \llap{1}3.10(0.10)            &
 \llap{1}1.45(0.05)            &
 \llap{1}0.25(0.05)            &
         9.00(0.05)            &
                               &
                               &
                               &
                               &
                               &
 CAM-CVF                       \\
SHV0500233$-$682914            &
 229                           &
 \llap{1}3.05(0.05)            &
 \llap{1}1.30(0.05)            &
 \llap{1}0.00(0.05)            &
         8.50(0.05)            &
           0.076(0.001)        &
                               &
           0.004(0.033)        &
 \llap{$-$}0.004(0.097)        &
                               &
                               \\
                               &
 729                           &
 \llap{1}3.30(0.10)            &
 \llap{1}1.40(0.10)            &
 \llap{1}0.20(0.05)            &
         8.60(0.05)            &
                               &
                               &
                               &
                               &
                               &
 CAM-CVF                       \\
SHV0502469$-$692418            &
 304                           &
 \llap{1}2.93(0.03)            &
 \llap{1}1.61(0.02)            &
 \llap{1}0.81(0.02)            &
 \llap{1}0.30(0.05)            &
           0.002(0.001)        &
                               &
                               &
                               &
                               &
                               \\
SHV0521050$-$690415            &
 788                           &
 \llap{1}1.07(0.02)            &
         9.77(0.02)            &
         9.23(0.02)            &
         8.40(0.10)            &
           0.062(0.004)        &
                               &
 \llap{$-$}0.011(0.014)        &
                               &
                               &
                               \\
SHV0522023$-$701242            &
 788                           &
 \llap{1}2.60(0.10)            &
 \llap{1}1.60(0.10)            &
 \llap{1}1.40(0.10)            &
                               &
           0.001(0.001)        &
                               &
           0.003(0.021)        &
                               &
                               &
                               \\
SHV0522118$-$702517            &
 217                           &
 \llap{1}3.00(0.20)            &
 \llap{1}2.00(0.20)            &
 \llap{1}1.00(0.20)            &
         9.70(0.20)            &
           0.028(0.001)        &
                               &
           0.007(0.002)        &
           0.170(0.125)        &
                               &
                               \\
SHV0524565$-$694559            &
 217                           &
 \llap{1}2.50(0.04)            &
 \llap{1}1.40(0.05)            &
 \llap{1}0.75(0.05)            &
                               &
           0.003(0.001)        &
                               &
                               &
                               &
                               &
                               \\
SHV0526001$-$701142            &
 217                           &
 \llap{1}3.50(0.10)            &
 \llap{1}2.00(0.10)            &
 \llap{1}0.45(0.05)            &
         9.15(0.05)            &
           0.037(0.001)        &
                               &
 \llap{$-$}0.019(0.030)        &
           0.051(0.063)        &
                               &
                               \\
SHV0530323$-$702216            &
 788                           &
 \llap{1}1.80(0.10)            &
 \llap{1}0.80(0.10)            &
 \llap{1}0.35(0.05)            &
                               &
           0.008(0.002)        &
                               &
           0.008(0.009)        &
                               &
                               &
                               \\
SHV0535442$-$702433            &
 782                           &
 \llap{1}2.90(0.20)            &
 \llap{1}1.50(0.10)            &
 \llap{1}0.65(0.05)            &
 \llap{1}0.30(0.20)            &
           0.009(0.001)        &
                               &
           0.029(0.022)        &
                               &
                               &
                               \\
SP77 30$-$6                    &
 195                           &
 \llap{1}0.80(0.10)            &
         9.60(0.10)            &
         9.20(0.10)            &
         8.50(0.10)            &
           0.139(0.001)        &
                               &
           0.077(0.018)        &
           0.225(0.073)        &
                               &
                               \\
                               &
 794                           &
 \llap{1}0.30(0.10)            &
         9.30(0.10)            &
         8.80(0.10)            &
         8.20(0.10)            &
                               &
                               &
                               &
                               &
                               &
 CAM-CVF                       \\
TRM45                          &
 788                           &
 \llap{1}6.20(0.20)            &
 \llap{1}3.60(0.10)            &
 \llap{1}1.55(0.05)            &
         9.50(0.10)            &
           0.076(0.001)        &
                               &
           0.037(0.007)        &
                               &
                               &
                               \\
TRM72                          &
 788                           &
 \llap{1}5.22(0.20)            &
 \llap{1}2.81(0.03)            &
 \llap{1}0.96(0.02)            &
         8.60(0.05)            &
           0.161(0.001)        &
                               &
           0.026(0.025)        &
                               &
                               &
                               \\
TRM88                          &
 788                           &
 \llap{1}3.50(0.10)            &
 \llap{1}1.78(0.02)            &
 \llap{1}0.24(0.02)            &
         8.55(0.05)            &
           0.101(0.001)        &
                               &
           0.051(0.014)        &
                               &
                               &
                               \\
WBP14                          &
 209                           &
 \llap{1}3.35(0.05)            &
 \llap{1}1.60(0.05)            &
 \llap{1}0.55(0.03)            &
         9.50(0.10)            &
           0.022(0.001)        &
                               &
 \llap{$-$}0.005(0.011)        &
           1.372(0.539)        &
                               &
                               \\
                               &
 616                           &
 \llap{1}3.40(0.10)            &
 \llap{1}1.60(0.10)            &
 \llap{1}0.55(0.05)            &
         9.50(0.05)            &
                               &
                               &
                               &
                               &
           0.211(0.293)        &
                               \\
WOH G64                        &
 229                           &
         9.58(0.02)            &
         7.96(0.02)            &
         6.98(0.02)            &
         5.32(0.02)            &
                               &
   \llap{1}2.112(0.965)        &
   \llap{1}3.786(0.354)        &
           4.647(0.298)        &
                               &
 PHOT-S                        \\
WOH SG374                      &
 788                           &
         9.90(0.02)            &
         9.10(0.02)            &
         8.64(0.02)            &
         7.69(0.05)            &
           0.490(0.005)        &
                               &
           0.188(0.020)        &
                               &
                               &
                               \\
\hline
\end{tabular}
\end{table}
\end{landscape}

%
% FIGURE 3
%
\begin{figure}[tb]
\centerline{\psfig{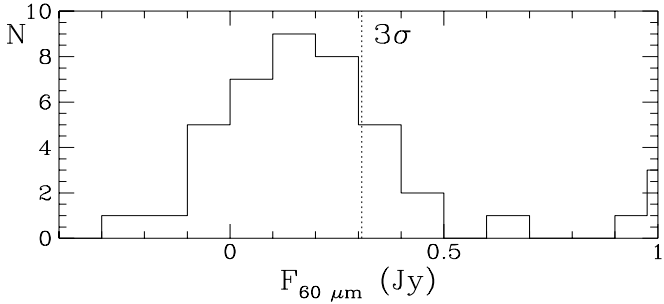}}
\caption[]{Histogram of the distribution of ISO 60 $\mu$m flux densities
(chopped measurements). The dotted vertical line indicates 3-$\sigma$ flux
density derived from the distribution of negative flux densities. All flux
densities over 1 Jy are piled up in the last bin.}
\end{figure}

A histogram of the distribution of ISO 60 $\mu$m flux densities (Fig.\ 3),
leaving out the mapping observations, illustrates the detection rate.
Considering negative flux densities indicating non-detection, and assuming a
Gaussian distribution around zero flux density for non-detections, we estimate
a 1-$\sigma$ detection to have 0.103 Jy. There are 12 sources with flux
densities exceeding 3-$\sigma$, i.e.\ probable detections. This does not take
into the account the large errors on some of the individual measurements, and
a 3-$\sigma$ detection may still turn out to be spurious (an example is
WBP14). On the other hand, the distribution below 3-$\sigma$ is certainly
skewed towards positive flux densities. Projecting the negative flux density
distribution onto the positive domain, we estimate that there are probably 17
more detections between 0 and 3-$\sigma$, and a total of 14 non-detections.

In the IRAS 60 $\mu$m data we found 8 detections and 17 tentative detections
(Table 3). The 0.1 Jy assigned to the faintest IRAS 60 $\mu$m flux densities
compares well with the ISO 1-$\sigma$ detection threshold of the chopped
measurements. Of the 8 IRAS 60 $\mu$m detections 6 have ISO chopped
measurements, all of which yield higher flux densities than IRAS --- by a
factor 1.8 on average. This is in contrast to the 12 and 25 $\mu$m photometry,
where ISO flux densities are generally lower than those measured by IRAS. The
14 ISO chopped measurements of IRAS tentative detections also yield higher
flux densities than did IRAS --- by a factor of 1.5 on average, although some
individual ISO measurements are fainter than the IRAS ones. None of these ISO
measurements is negative, indicating that many of the IRAS 60 $\mu$m tentative
detections are indeed real.

The 7 mapping observations all agree with the chopped measurements within
2-$\sigma$, although these errors can be large. There is no tendency for one
of these two methods to yield higher flux densities than the other. As we do
not expect strong variability at 60 $\mu$m, which traces cool dust some
distance from the stars, flux densities from mapping and chopped measurements
are averaged. The error estimates of the mapping measurements are
systematically larger than those of the chopped measurements. This may be due
to the fact that, for the mapping data, the flux density of the star was
determined from the inner $3\times3$ pixels. The contribution of the
background to these 9 pixels is considerable. Also, the reliability of the
error estimate for the central pixel in the chopped data as produced by PIA is
unknown. There is great difficulty in extracting reliable photometry and
associated errors from either mapping or chopped measurements at 60 $\mu$m,
for stellar sources in fields like the LMC. This is mainly due to the complex
background and limited spatial resolution of PHOT at these wavelengths.
IRAS05289$-$6617 has a very smooth background, being situated in the
line-of-sight to the supergiant shell LMC4 (Meaburn 1980). Indeed, ISO mapping
and chopped measurements are relatively precise for this source, and agree
nicely with the 60 $\mu$m flux density measured from the IRAS data.

%
% FIGURE 4
%
\begin{figure}[tb]
\centerline{\psfig{figure=h1374.f4,width=88mm}}
\caption[]{The CAM-CVF spectra of obscured AGB stars in the LMC. Open symbols
represent spectro-photometric points that are prone to have flux densities
that are over-estimated due to stabilisation difficulties. The spectral shape
is best represented by the solid symbols (squares for the short-, disks for
the long-wavelength region). Emission and/or absorption centred at $\sim9.7$
$\mu$m is indicative of oxygen-rich dust (e.g.\ IRAS05402$-$6956 and SP77
30$-$6), whereas carbon-rich dust may show emission at $\sim11.3$ $\mu$m
(e.g.\ IRAS05289$-$6617). A featureless continuum around 10 $\mu$m also
strongly suggests carbon-rich dust (e.g.\ SHV0500193$-$681706).}
\end{figure}

%
% FIGURE 5a
%
\begin{figure*}[tb]
\centerline{\psfig{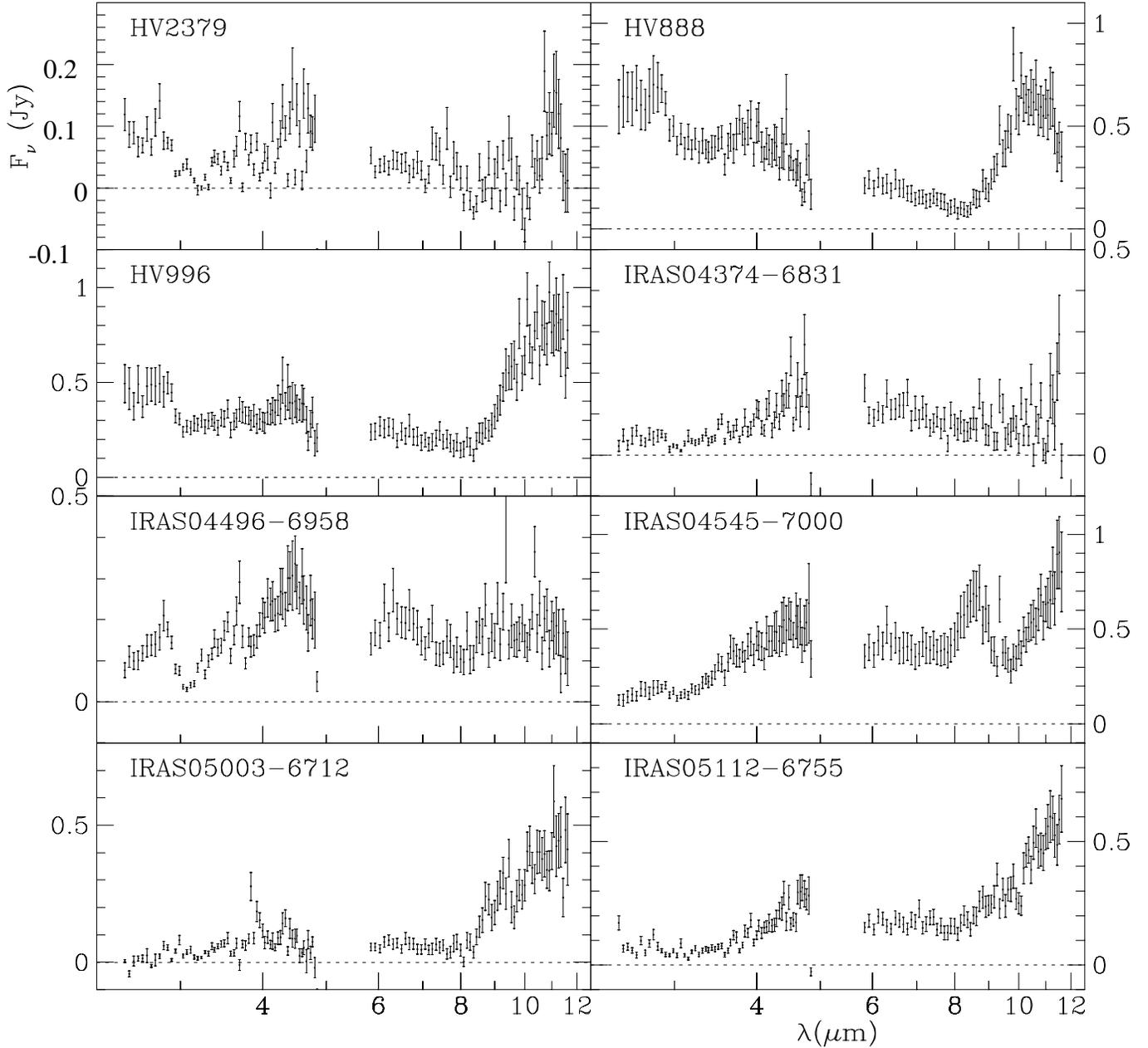}}
\caption[]{The PHOT-S spectra of obscured AGB stars (and RSGs) in the LMC.
Emission and/or absorption centred at $\sim9.7$ $\mu$m suggests oxygen-rich
dust (e.g.\ HV888 and IRAS04545$-$7000). Absorption at 3 $\mu$m is seen in
carbon star photospheres (e.g.\ IRAS04496$-$6958), but artifacts in the PHOT-S
responsivities also mimic weak depression at 3 $\mu$m in the spectra of
unambiguous oxygen-rich stars (e.g.\ IRAS04545$-$7000).}
\end{figure*}

%
% FIGURE 5b
%
\begin{figure*}[tb]
\addtocounter{figure}{-1}
\centerline{\psfig{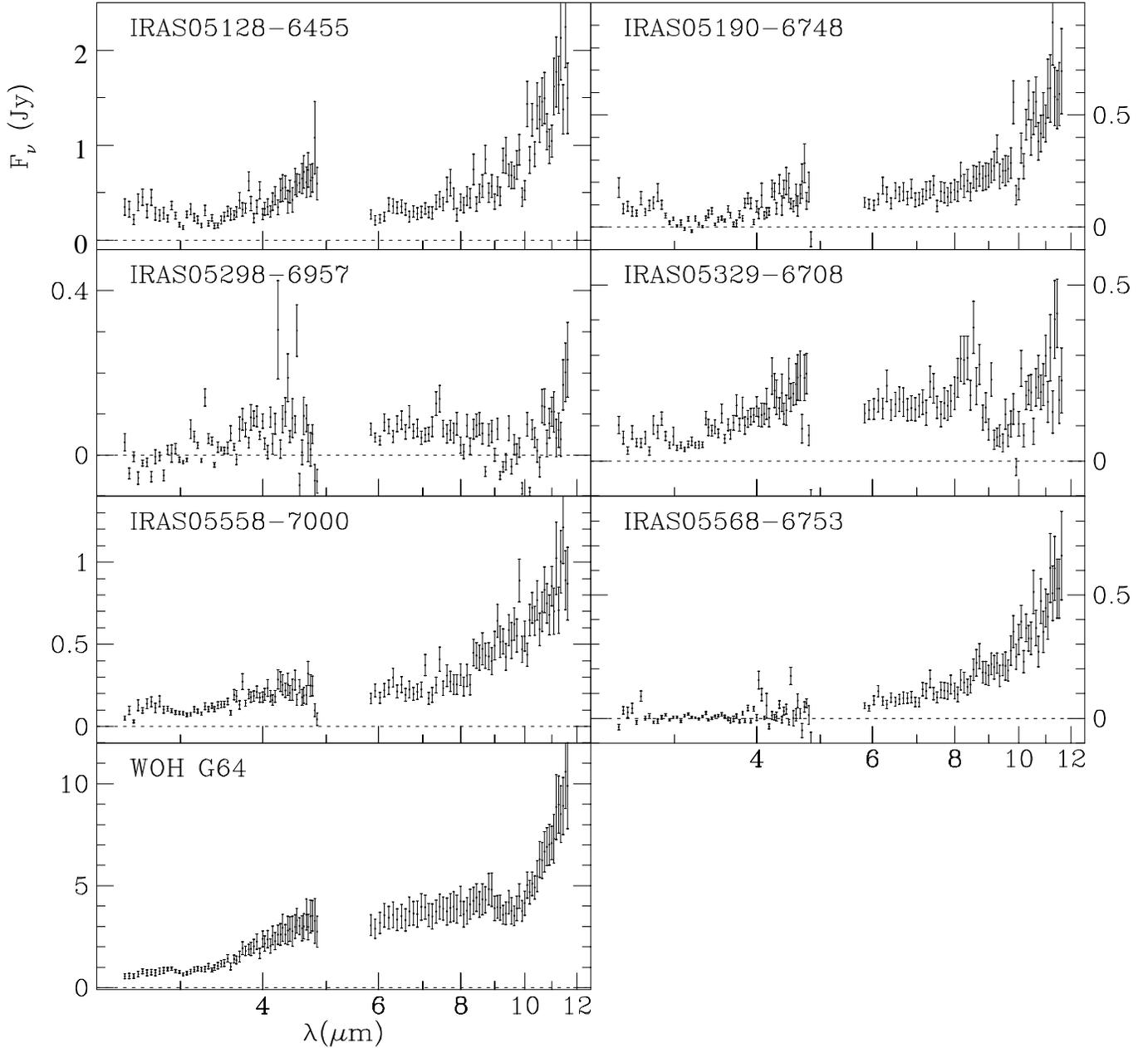}}
\caption[]{(continued)}
\end{figure*}

\section{Discussion}

\subsection{Chemical types from ISO spectroscopy}

The presence or absence in the ISO spectra (Figs.\ 4 \& 5) of discrete dust
emission and molecular absorption bands can be used to distinguish between
carbon- and oxygen-rich circumstellar envelopes (e.g.\ Merrill \& Stein
1976a,b,c). The results are summarised in Table 5.

Amorphous oxygen-rich dust may give rise to strong and broad silicate emission
between $\sim8$ and 13 $\mu$m, peaking at $\sim9.7 \mu$m (the exact location
may differ from this by a few tenths of $\mu$m). The late-M stars HV2446, 888,
996, and SP77 30$-$6 have prominent silicate emission. In optically thick
cases the silicate feature turns into absorption. All spectra of OH maser
sources show the silicate feature in self-absorption: IRAS04545$-$7000,
05298$-$6957, 05329$-$6708, 05402$-$6956, and WOH G64.

Oxygen-rich molecules do not provide clear diagnostics of the chemical type of
CSEs at our signal-to-noise and spectral resolution. We already mentioned that
shallow absorption around 3 $\mu$m in oxygen-rich sources is most likely due
to an artifact in the responsivities, rather than H$_2$O ice.

%
% TABLE 5a
%
\begin{table}
\caption[]{Chemical types. Optical spectra (Opt Sp) include objective prism
and CCD spectroscopy up to $\sim1 \mu$m. ISO spectroscopy (ISO Sp) comprises
PHOT-S and CAM-CVF observations. IR colour-colour diagrams (IR col) can in
some cases be reasonably conclusive too: we here use $(K-[12])$ and
$([12]-[25])$ versus $(K-L)$ diagrams. At radio wavelengths, OH, SiO and/or
H$_2$O maser emission is detected from some oxygen-rich sources.}
\begin{tabular}{lllll}
\hline\hline
Star                           &
Opt Sp                         &
ISO Sp                         &
IR col                         &
Maser                          \\
\hline
GRV0519$-$6700                 &
C                              &
                               &
carbon?                        &
                               \\
HV12070                        &
MS3/9                          &
oxygen?                        &
?                              &
                               \\
HV12501                        &
M1.5                           &
                               &
oxygen                         &
                               \\
HV2360                         &
M2 Ia                          &
                               &
oxygen                         &
                               \\
HV2379                         &
C                              &
SiC?                           &
carbon                         &
                               \\
HV2446                         &
M5e                            &
silicate                       &
oxygen                         &
                               \\
HV5870                         &
M4.5/5                         &
                               &
oxygen                         &
                               \\
HV888                          &
M4 Ia                          &
silicate                       &
oxygen                         &
                               \\
HV916                          &
M3 Iab                         &
                               &
oxygen                         &
                               \\
HV996                          &
M4 Iab                         &
silicate                       &
oxygen                         &
                               \\
IRAS04286$-$6937               &
                               &
                               &
carbon                         &
                               \\
IRAS04374$-$6831               &
                               &
SiC?                           &
carbon                         &
                               \\
IRAS04407$-$7000               &
                               &
                               &
oxygen                         &
yes                            \\
IRAS04496$-$6958               &
C                              &
car$+$sil?                     &
carbon                         &
                               \\
IRAS04498$-$6842               &
                               &
                               &
oxygen                         &
                               \\
IRAS04509$-$6922               &
M10                            &
                               &
oxygen                         &
                               \\
IRAS04516$-$6902               &
M9                             &
                               &
oxygen                         &
                               \\
IRAS04530$-$6916               &
                               &
                               &
oxygen                         &
                               \\
IRAS04539$-$6821               &
                               &
                               &
carbon                         &
                               \\
IRAS04545$-$7000               &
                               &
silicate                       &
oxygen                         &
yes                            \\
IRAS04557$-$6753               &
                               &
                               &
carbon                         &
                               \\
IRAS05003$-$6712               &
                               &
silicate?                      &
oxygen                         &
                               \\
IRAS05009$-$6616               &
                               &
                               &
carbon                         &
                               \\
IRAS05112$-$6755               &
                               &
carbon?                        &
carbon                         &
                               \\
IRAS05113$-$6739               &
                               &
                               &
carbon                         &
                               \\
IRAS05128$-$6455               &
                               &
carbon?                        &
carbon                         &
                               \\
IRAS05190$-$6748               &
                               &
carbon?                        &
carbon                         &
                               \\
IRAS05289$-$6617               &
                               &
SiC                            &
?                              &
                               \\
IRAS05294$-$7104               &
                               &
                               &
oxygen                         &
                               \\
IRAS05295$-$7121               &
                               &
                               &
carbon                         &
                               \\
IRAS05298$-$6957               &
                               &
silicate                       &
oxygen                         &
yes                            \\
IRAS05300$-$6651               &
                               &
                               &
carbon                         &
                               \\
IRAS05329$-$6708               &
                               &
silicate                       &
oxygen                         &
yes                            \\
IRAS05348$-$7024               &
                               &
SiC                            &
carbon                         &
                               \\
IRAS05360$-$6648               &
                               &
                               &
carbon                         &
                               \\
IRAS05402$-$6956               &
                               &
silicate                       &
oxygen                         &
yes                            \\
IRAS05506$-$7053               &
                               &
                               &
oxygen                         &
                               \\
IRAS05558$-$7000               &
                               &
silicate                       &
oxygen                         &
                               \\
IRAS05568$-$6753               &
                               &
carbon?                        &
?                              &
                               \\
SHV0454030$-$675031            &
C                              &
                               &
carbon                         &
                               \\
SHV0500193$-$681706            &
                               &
carbon                         &
carbon                         &
                               \\
SHV0500233$-$682914            &
                               &
SiC?                           &
carbon                         &
                               \\
SHV0502469$-$692418            &
C                              &
                               &
carbon?                        &
                               \\
SHV0521050$-$690415            &
C                              &
                               &
carbon                         &
                               \\
SHV0522023$-$701242            &
M3                             &
                               &
?                              &
                               \\
SHV0522118$-$702517            &
S?                             &
                               &
carbon                         &
                               \\
SHV0524565$-$694559            &
MS5                            &
                               &
?                              &
                               \\
SHV0526001$-$701142            &
C                              &
                               &
carbon                         &
                               \\
SHV0530323$-$702216            &
M6                             &
                               &
oxygen                         &
                               \\
SHV0535442$-$702433            &
C                              &
                               &
?                              &
                               \\
SP77 30$-$6                    &
M8                             &
silicate                       &
oxygen                         &
                               \\
TRM45                          &
                               &
                               &
carbon                         &
                               \\
\hline
\end{tabular}
\end{table}

%
% TABLE 5b
%
\begin{table}
\addtocounter{table}{-1}
\caption[]{(continued).}
\begin{tabular}{lllll}
\hline\hline
Star                           &
Opt Sp                         &
ISO Sp                         &
IR col                         &
Maser                          \\
\hline
TRM72                          &
C                              &
                               &
carbon                         &
                               \\
TRM88                          &
C                              &
                               &
carbon                         &
                               \\
WBP14                          &
C                              &
                               &
carbon                         &
                               \\
WOH G64                        &
M7.5                           &
silicate                       &
oxygen                         &
yes                            \\
WOH SG374                      &
M6                             &
                               &
oxygen                         &
                               \\
\hline
\end{tabular}
\end{table}

Crystalline carbon-rich dust sometimes gives rise to a SiC (graphite) emission
feature peaking at $\sim11.3 \mu$m, and narrower than the silicate feature.
The CVF spectrum of IRAS05289$-$6617 (Fig.\ 4) shows the best example of this.

Carbon-rich molecules have several strong absorption bands in our spectral
region, all from HCN and C$_2$H$_2$. The strongest is at 3.1 $\mu$m, but the
problem with the responsivities limits the number of unambiguous detections to
one (IRAS04496$-$6958). Related, but weaker, absorption is visible at 3.8
$\mu$m. More absorption bands are located around 5, 8 and 14 $\mu$m.
Unfortunately, the 5 $\mu$m band falls entirely in the blind spectral region
of PHOT-S. The 8 and 14 $\mu$m bands are at the edges of the CVF spectra and
hence difficult to identify.

Some other spectra show merely a featureless dust continuum around 10 $\mu$m.
Best examples are the CVF spectrum of SHV0500193$-$681706 and the PHOT-S
spectrum of IRAS05568$-$6753. These spectra suggest pure amorphous carbon dust
emission.

\subsection{IR colour-colour diagrams}

The ISO 12, 25 and 60 $\mu$m filters are similar but not identical to the IRAS
filters. As the zero-points of these ISO filters are unknown, we adopt here
the IRAS zero-points. This results in the following definitions for the (not
colour-corrected) mid-IR magnitudes
\begin{equation}
[12]=-2.5\log(F_{12}/28.3)
\end{equation}
\begin{equation}
[25]=-2.5\log(F_{25}/6.73)
\end{equation}
\begin{equation}
[60]=-2.5\log(F_{60}/1.19)
\end{equation}

\subsubsection{Diagram of $(K-[12])$ versus $(H-K)$}

%
% FIGURE 6
%
\begin{figure}[tb]
\centerline{\psfig{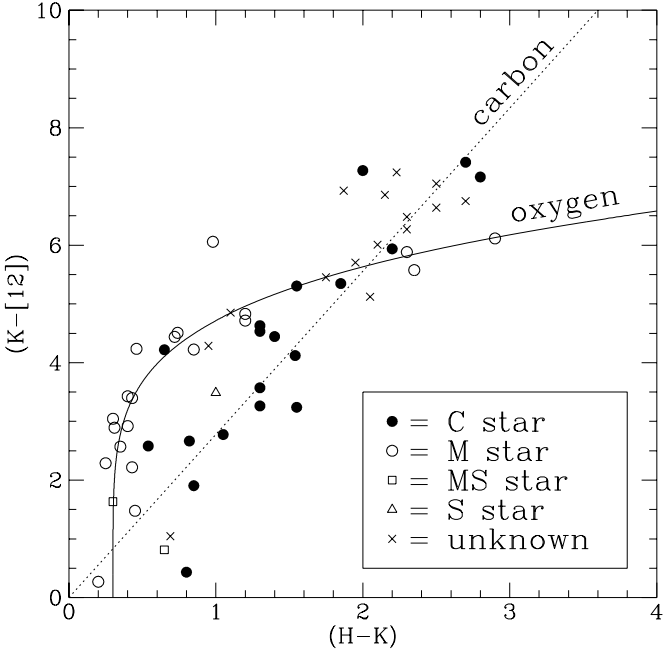}}
\caption[]{$(K-[12])$ versus $(H-K)$ diagram. Stars are distinguished by their
chemical types inferred from spectroscopic methods: carbon stars (solid
disks), M stars (open disks), MS stars (open squares), S stars (open
triangles), and stars of which the chemical type is a priori unknown
(crosses). Carbon stars and oxygen stars define sequences in this diagram,
indicated by a dotted and solid curve, respectively.}
\end{figure}

The $(K-[12])$ versus $(H-K)$ colour-colour diagram separates carbon- from
oxygen-rich stars in samples of obscured stars in the MCs (Papers II \& IV).
Indeed, the distributions of carbon- and oxygen-rich stars using ISO and SAAO
photometry define clear sequences in this diagram (Fig.\ 6). The sequences are
fit by eye, with the carbon sequence the same as in Paper IV:
\begin{equation}
(H-K)=0.36\times(K-[12])
\end{equation}
but the oxygen sequence a simple, yet somewhat steeper function than in Paper
IV:
\begin{equation}
(H-K)=0.3+0.0003\times(K-[12])^5
\end{equation}
Although the stars with spectral type M follow the oxygen sequence very well,
the carbon stars show a large scatter around the carbon sequence with several
carbon stars on or beyond the region populated by M stars, at small $(H-K)$
but large $(K-[12])$ magnitudes. This scatter contrasts with the tight carbon
sequence that is observed in the Milky Way (Fig.\ 3 in Paper IV). We suspect
that this is in part caused by the severe crowding in some LMC fields,
affecting the near-IR aperture photometry. Differences in the strength of
absorption in the H-band by carbonaceous molecules may cause additional
scatter among carbon stars (Bessell \& Wood 1983; Catchpole \& Whitelock
1985).

\subsubsection{Diagram of $(K-[12])$ versus $(K-L)$}

The $(K-[12])$ versus $(K-L)$ colour-colour diagram shows much less scatter
around well-defined carbon and oxygen sequences (Fig.\ 7). This makes it a
much more powerful diagnostic diagram than the $(K-[12])$ versus $(H-K)$
diagram in typifying the chemical composition of the circumstellar dust.
Noguchi et al.\ (1991a) introduced a very similar diagnostic using $(L-[12])$
and $(K-L)$ colours. We note, however, that some of the peculiar stars in our
$(K-[12])$ versus $(H-K)$ diagram were too blue and hence too faint to be
detected in the L-band. Still, the tight sequences prove that both the SAAO
and ISO photometry are of good quality when comparing individual stars. We fit
(by eye) a linear carbon sequence:
\begin{equation}
(K-L)=\frac{5}{11}\times(K-[12])-\frac{2}{11}
\end{equation}
and a superposition of even polynomials for the oxygen sequence:
\begin{equation}
(K-L)=0.35+0.007\times(K-[12])^2+0.0014\times(K-[12])^4
\end{equation}

%
% FIGURE 7
%
\begin{figure}[tb]
\centerline{\psfig{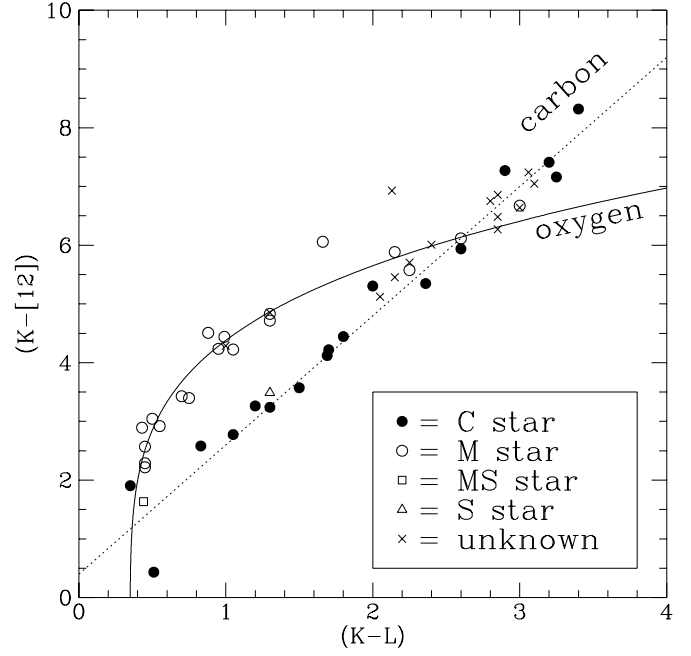}}
\caption[]{$(K-[12])$ versus $(K-L)$ diagram. Symbols as in Fig.\ 6.}
\end{figure}

\subsubsection{Diagram of $([12]-[25])$ versus $(K-L)$}

%
% FIGURE 8
%
\begin{figure}[tb]
\centerline{\psfig{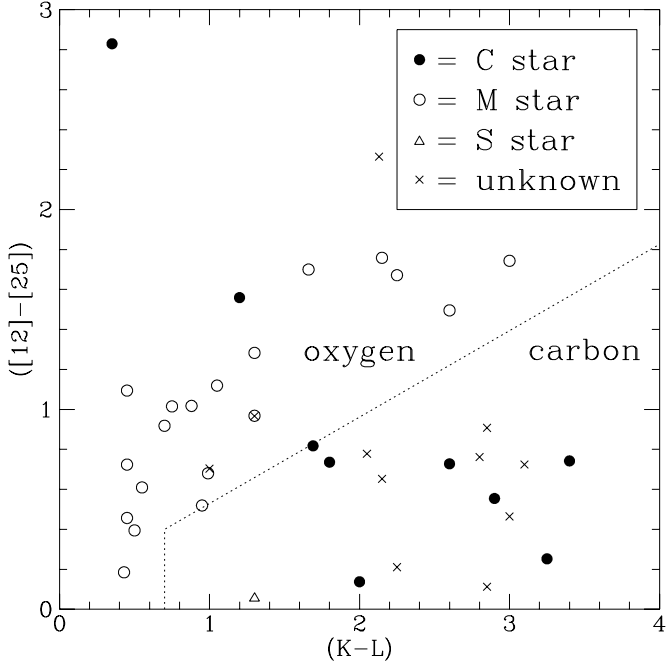}}
\caption[]{$([12]-[25])$ versus $(K-L)$ diagram. Symbols are as in Fig.\ 6.
Oxygen-rich sources and carbon stars occupy distinct areas in this diagram.
The dividing line (dotted) between stars with carbon- and oxygen-rich dust is
taken from Epchtein et al.\ 1987.}
\end{figure}

Another colour-colour diagram that separates carbon- from oxygen-rich sources
is the $([12]-[25])$ versus $(K-L)$ diagram (Fig.\ 8). The confirmed
oxygen-rich sources show a linear relationship between the $([12]-[25])$ and
$(K-L)$ colours, possibly flattening out at $(K-L)>1.5$ mag. The LMC stars
generally follow the separation determined for galactic stars (dotted line in
Fig.\ 8, taken from Epchtein et al.\ 1987).

\subsubsection{Diagram of $([25]-[60])$ versus $([12]-[25])$}

%
% FIGURE 9
%
\begin{figure}[tb]
\centerline{\psfig{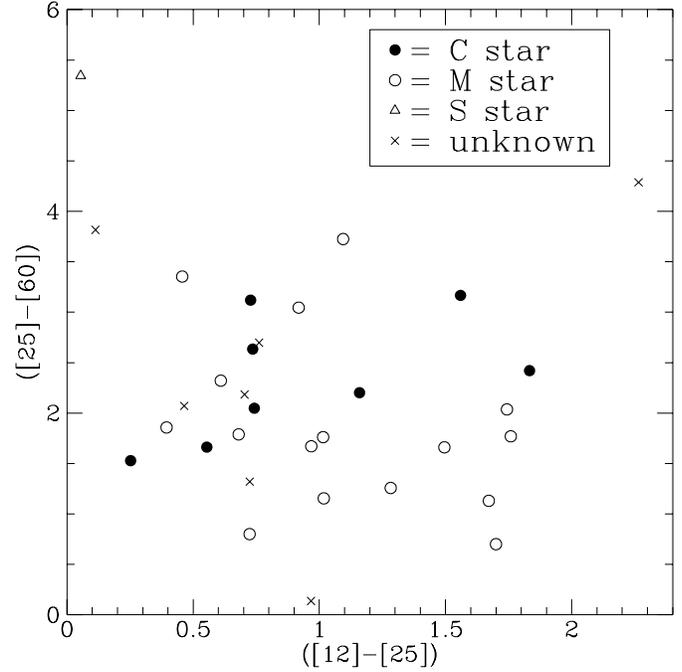}}
\caption[]{$([25]-[60])$ versus $([12]-[25])$ diagram. Symbols are as in Fig.\
6. Carbon stars are not well separated from oxygen-rich sources, although
carbon stars seem to be relatively bright at 60 $\mu$m.}
\end{figure}

The aim of obtaining 60 $\mu$m flux densities for stars in the LMC is mainly
to probe the coolest circumstellar dust. The 60 $\mu$m flux density is
expected to increase as prolonged mass loss first extends the CSE and again as
reduced mass loss results in a detached shell. This evolution might be seen in
$([25]-[60])$ versus $([12]-[25])$ diagrams (Fig.\ 9, see also van der Veen \&
Habing 1988). Unfortunately, the accuracy of the ISO photometry at 60 $\mu$m
is not very high for most of these LMC sources, and the diagram contains a lot
of scatter.

Perhaps the most obvious thing to learn from this diagram is that carbon stars
tend to be relatively bright at 60 $\mu$m, yielding $([25]-[60])\sim1.5$ to 3
mag. Although oxygen-rich sources can have similar colours, there are many
oxygen-rich sources with $([25]-[60])<2$ and $([12]-[25])>0.6$ mag, colours
not seen for any carbon star in our sample. This is similar to the findings of
van der Veen \& Habing (1988), but our LMC sources have bluer $([12]-[25])$
and redder $([25]-[60])$ colours than do their Milky Way sources. However, the
LMC $([12]-[25])$ colours do not differ much from those discussed by Le Bertre
et al.\ (1994).

\subsection{Comments on particular objects}

\subsubsection{GRV0519$-$6700}

The referee Dr.\ Peter Wood conveys that an optical spectrum of GRV0519$-$6700
shows it to be a carbon star, in good agreement with its IR colours of
$(H-K)=0.7$ and $(K-[12])=1.0$ mag.

\subsubsection{HV12070}

The CVF spectrum of HV12070 shows only a hint of the silicate feature, whilst
the IR colours cannot distinguish between oxygen- and carbon-rich dust of the
optically thin CSE of this MS-type star.

\subsubsection{HV2379}

The PHOT-S spectrum of HV2379 suggests SiC emission, but its CVF spectrum does
not. This may be a result of changes in the properties of the CSE or the dust.
Its IR colours leave no doubt about the carbon-rich nature of the dust.

\subsubsection{HV2446, 888, 996, and SP77 30$-$6}

These late-M stars all have prominent silicate emission and IR colours that
unambiguously indicate oxygen-rich dust.

\subsubsection{IRAS04286$-$6937, 04539$-$6821, 04557$-$6753, 05009$-$6616,
05113$-$6739, 05295$-$7121, 05300$-$6651, 05360$-$6648, and TRM45 and 72}

The position of these objects in the $(K-[12])$ versus $(H-K)$ or $(K-L)$
colour-colour diagrams does not clarify the chemical composition of their
CSEs. The $([12]-[25])$ versus $(K-L)$ diagram, however, unambiguously
indicates that the dust around these stars is carbon rich. The IR colours of
IRAS05113$-$6739 at the three ISO epochs for this star all lie along the
carbon sequences in the $(K-[12])$ versus $(H-K)$ and $(K-L)$ diagrams.
Ground-based L-band spectra of IRAS05009$-$6616 and 05300$-$6651 show the 3.1
$\mu$m absorption feature due to HCN and C$_2$H$_2$ molecules, indicating
carbon-rich photospheres (van Loon et al.\ 1999).

\subsubsection{IRAS04374$-$6831}

The position of IRAS04374$-$6831 in the $([12]-[25])$ versus $(K-L)$ diagram
indicates carbon-rich dust. Its PHOT-S spectrum, which does not clearly reveal
the chemical composition of the dust by itself, is then marginally consistent
with SiC emission.

\subsubsection{IRAS04496$-$6958}

IRAS04496$-$6958 shows strong absorption by carbonaceous molecules at 3.1
$\mu$m, already known from ground-based L-band spectroscopy (van Loon et al.\
1999). Related, but weaker, absorption is visible at 3.8 $\mu$m, and possibly
around 8 $\mu$m. Surprisingly, this carbon star has silicate emission too,
indicating the presence of oxygen-rich dust (see Trams et al.\ 1999). Its IR
colours indicate carbon-rich dust, hence the oxygen-rich dust is only a minor
component.

\subsubsection{IRAS04530$-$6916}

With $(K-L)=2.13$, $(K-[12])=6.9$ and $([12]-[25])=2.3$ mag, the IR colours of
IRAS04530$-$6916 imply that the dust around this very luminous and red object
must be oxygen rich.

\subsubsection{IRAS04545$-$7000, 05298$-$6957, 05329$-$6708, 05402$-$6956, and
WOH G64}

These OH maser sources all show the silicate feature in self-absorption, and
also their IR colours clearly indicate oxygen-rich dust.

\subsubsection{IRAS05003$-$6712}

The IR colours of IRAS05003$-$6712 unambiguously classify the dust as oxygen
rich. The PHOT-S spectrum shows a hint of the silicate feature. A ground-based
L-band spectrum of this star shows a featureless continuum around 3.1 $\mu$m,
indicating an oxygen-rich photosphere (van Loon et al.\ 1999).

\subsubsection{IRAS05112$-$6755}

The dust around IRAS05112$-$6755 is classified as carbon rich on the basis of
the position in the $([12]-[25])$ versus $(K-L)$ diagram. There is a hint of 8
$\mu$m absorption in the PHOT-S spectrum of IRAS05112$-$6755. A ground-based
L-band spectrum of this object shows the strong absorption at 3.1 $\mu$m found
in carbon-rich stellar photospheres (van Loon et al.\ 1999).

\subsubsection{IRAS05128$-$6455 and 05190$-$6748}

The absence of clear indications for the presence of the silicate feature in
the PHOT-S spectra of these stars suggest that their dust may be carbon rich,
which is also indicated by their $([12]-[25])$ and $(K-L)$ colours.

\subsubsection{IRAS05289$-$6617}

The CVF spectrum of IRAS05289$-$6617 shows prominent SiC emission. Hence it is
probably a mass-losing carbon-rich AGB star in the LMC rather than a
foreground object. We have not yet identified its near-IR counterpart.

\subsubsection{IRAS05348$-$7024}

The CVF spectrum of IRAS05348$-$7024 shows weak SiC emission. The carbon-rich
nature of the dust around this object is also indicated by its position in the
$([12]-[25])$ versus $(K-L)$ diagram.

\subsubsection{IRAS05506$-$7053}

IRAS05506$-$7053 is the only star in our sample that could not be detected at
12 $\mu$m. Assuming a 12 $\mu$m flux density $<0.03$ Jy, the $(K-[12])$ colour
would be $<6.2$ mag and probably $([12]-[25])>1.5$ mag. At $(K-L)=3.3$ mag,
this suggests an oxygen-rich CSE.

\subsubsection{IRAS05558$-$7000}

The CVF spectrum of IRAS05558$-$7000 is similar to the CVF spectra of
IRAS05298$-$6957 and 05402$-$6956, showing silicate emission that is becoming
optically thick at 10 $\mu$m. The IR colours of IRAS05558$-$7000 unambiguously
imply that the dust is oxygen-rich.

\subsubsection{IRAS05568$-$6753}

The PHOT-S spectrum of IRAS05568$-$6753 shows a featureless dust continuum
around 10 $\mu$m, suggesting pure amorphous carbon dust emission. The near-IR
counterpart of this object has yet to be identified.

\subsubsection{SHV0500193$-$681706}

The CVF spectrum of SHV0500193$-$681706 shows a featureless dust continuum
around 10 $\mu$m, suggesting pure amorphous carbon dust emission. The
carbon-rich nature of the dust is confirmed by the position in the $(K-[12])$
versus $(K-L)$ diagram. Inaccuracy of its 25 $\mu$m flux density causes the
rather odd position among the oxygen-rich stars in the $([12]-[25])$ versus
$(K-L)$ diagram.

\subsubsection{SHV0500233$-$682914}

The CVF spectrum of SHV0500233$-$682914 shows a hint of SiC emission, and also
its IR colours clearly indicate that the dust around this star is carbon rich.

\subsubsection{SHV0502469$-$692418, 0522023$-$701242 and 0524565$-$694559}

The carbon star SHV0502469$-$692418, the M-type star SHV0522023$-$701242 and
the MS-type star SHV0524565$-$694559 are surrounded by an optically thin CSE
and hence it is difficult to classify the chemical type of their dust from IR
colour-colour diagrams.

\subsubsection{SHV0522118$-$702517}

SHV0522118$-$702517 was tentatively classified an S-type star by Hughes \&
Wood (1990). Its IR colours are clearly similar to those of carbon stars. This
suggests that carbon-rich dust dominates the absorption and emission
characteristics of the CSE despite the under-abundance of carbon atoms in its
photosphere. Noguchi et al.\ (1991b) show that the IR colours of the CSE
indicate oxygen-rich dust in case of an MS-type star. Also, CS stars show 3
$\mu$m absorption from HCN and C$_2$H$_2$ molecules, whereas SC stars do not
(Catchpole \& Whitelock 1985; Noguchi \& Akiba 1986). This suggests that
carbon chemistry is dominant in CS stars, but not in SC stars. Thus, we
identify SHV0522118$-$702517 with a CS star. Dust-enshrouded S stars ---
including MS and CS stars --- that have $(K-L)>1$ mag are very rare in the
Milky Way, and none are known with $(K-L)>2$ mag (Noguchi et al.\ 1991b).
Hence, with $(K-L)=1.3$ mag, SHV0522118$-$702517 is among the most obscured S
stars known.

\subsubsection{SHV0530323$-$702216}

The late-M type star SHV0535442$-$702433 has $([12]-[25])=1.56$ mag. Its
near-IR colours are rather blue and $(K-L)$ is not expected to be larger than
unity. Hence the position of this object in the $([12]-[25])$ versus $(K-L)$
diagram suggests that the dust is oxygen rich.

\subsubsection{SHV0535442$-$702433}

The carbon star SHV0535442$-$702433 is surrounded by an optically thin CSE,
and hence the IR colours are difficult to use for classifying the chemical
type of the dust. The location among oxygen-rich stars in the $([12]-[25])$
versus $(K-L)$ diagram is caused entirely by the inaccuracy of its 25 $\mu$m
flux density yielding a spuriously red $([12]-[25])\sim3$ mag.

\section{Conclusions}

ISO spectroscopy is used to determine the chemical type of the dust around
obscured cool evolved stars in the LMC. ISO photometry at 12, 25 and 60 $\mu$m
is presented, together with quasi-simultaneous near-IR photometry from the
ground (SAAO). The accuracy and sensitivity of the ISOPHOT photometry is not
much better than can be achieved from properly treated IRAS data. The ISOCAM
photometry is much more reliable because it is based on imaging, and an order
of magnitude more sensitive than was IRAS. Colour-colour diagrams prove that
relative photometry is reliable. A combination of $(K-[12])$ and $([12]-[25])$
versus $(K-L)$ diagrams provide a reliable way of distinguishing between
carbon- and oxygen-rich dust, provided the CSE has sufficient optical depth.
The combination of ISO spectra and photometry enabled us to securely classify
the chemical type of the dust around nearly all stars in our sample. This was
previously known for only a minority of the stars. Surprisingly, the
$(K-[12])$ versus $(H-K)$ diagnostic diagram contains a lot of scatter
especially among carbon stars.

Many of the obscured AGB stars in our sample are carbon stars: 46\% amongst
the LMC stars that were detected by IRAS (Table 1). M stars were always found
to be surrounded by oxygen-rich dust. In particular, all detected OH maser
sources show self-absorbed silicate emission. As in the Milky Way, the fact
that no M star with carbon-rich dust has ever been found suggests that HBB
cannot efficiently turn carbon stars back into oxygen-rich stars. The dust
around the dust-enshrouded S star SHV0522118$-$702517 has the characteristics
of carbon-rich material, suggesting it is actually a CS star. Surprisingly,
the dust around the carbon star IRAS04496$-$6958 has a (minor) oxygen-rich
component (Trams et al.\ 1999).

\acknowledgements{We would like to thank everyone at VilSpa (Madrid) for the
discussions and advices during the various stages of ISO data reduction, in
particular Drs.\ Jos\'{e} Acosta-Pulido, Carlos Gabriel, Rene Laureijs, Thomas
M\"{u}ller, and Bernhard Schulz, and Dr.\ P\'{e}ter \'{A}brah\'{a}m in
Heidelberg. The ISOCAM data presented in this paper was analysed using
``CIA'', a joint development by the ESA Astrophysics Division and the ISOCAM
Consortium. The ISOCAM Consortium is led by the ISOCAM PI, C. Cesarsky,
Direction des Sciences de la Mati\'{e}re, C.E.A., France. The ISOPHOT data
presented in this paper was reduced using PIA, which is a joint development by
the ESA Astrophysics Division and the ISOPHOT Consortium. We also thank Dr.\
Romke Bontekoe for help with studying the IRAS data, and the referee Dr.\
Peter Wood for his suggestions that improved the presentation considerably. We
made use of the SIMBAD database, operated at CDS, Strasbourg, France. This
research was partly supported by NWO under Pionier Grant 600-78-333.\\
(JvL: O trabalho mais importante foi feito por um anjo).}

\appendix

\section{ISO photometry}

\subsection{CAM 12 $\mu$m imaging-photometry}

The ISOCAM images were corrected for the dark image valid for the
corresponding revolution and orbital position of the spacecraft, and corrected
for glitches using the multiresolution median transform method. Although the
current dark subtraction algorithm produces satisfactory results, we
nonetheless also applied our own IDL routine that optimises the dark
subtraction by appropriately scaling the dark and flatfield images in a
rectangular annulus leaving a $12\times12$ pixel area centred at the stellar
position on the array. The dark subtraction was improved slightly in some
cases, whereas for the majority of the observations there was no significant
difference. For each pixel, we applied a $\kappa$-$\sigma$ rejection criterion
to the evolution of the signal in time: values that deviate more than 2
$\sigma$ from the time-average were rejected. The images were divided by the
default calibration flatfield image to correct for the array pattern in the
pixel responsivities, and time-averaged.

Point source photometry was performed on the final image, using our own
procedures written within the Munich Interactive Data Analysis Software
(MIDAS). We measured the flux density within a circular area (software
aperture) around the stellar position, and subtracted the background flux
density level determined from a concentric circular annulus between 12 and 14
pixels radius from the stellar position --- excluding the pixels near the edge
of the $32\times32$ pixels array where vignetting is evident. We repeated this
for different software aperture radii, to create a magnitude profile (MP) of
the stellar flux density versus radius. We used the 23 brightest stars ---
with flux densities ranging from 0.06 to 0.8 Jy --- to create a template MP.
Comparison of the individual MPs of the stars with the template yielded
differential MPs. Where the differential MP is constant with radius, reliable
differential magnitudes can be determined. We adopt the standard deviation in
these points as 1-$\sigma$ errors on the differential photometry. The strength
of the MP-method is the estimation of photometric errors, as well as the
selection of the part of the MP that best resembles the PSF. Calibration of
the template MP was done using the calibration conversion 4.13 ADU s$^{-1}$
mJy$^{-1}$ (OLP V7 calibration). The template MP was consistent with a
synthetic MP created from the known PSF, and indicated an uncertainty of
$\sim1$ mJy.

All target stars were detected unambiguously, except for SHV0522023$-$70124
which was marginally detected. Most were well centred --- within about
$10^{\prime\prime}$ --- but IRAS05113$-$6739 (last epoch only),
IRAS05295$-$7121 and TRM45 were off by 20 to $30^{\prime\prime}$. Seven stars
were accompanied by generally weaker additional point sources within the
image: HV12501, SHV0500193$-$681706, 0500233$-$682914, 0502469$-$692418, and
0535442$-$702433, and TRM72 and 88. All of these stars are either HV, SHV, or
TRM sources, and the nature of the additional point sources remains a mystery.

\subsection{PHOT-P 12 and 25 $\mu$m chopped measurements}

The measurements each consist of 4 Destructive Readouts (DRs), with 127
Non-Destructive Readouts (NDRs) per ramp. All DRs and the first 8 NDRs per
ramp were discarded, because these read-outs are considered unreliable due to
the strong impact of the detector reset. The signal corresponding to the
source intensity is derived from the slope of the integration ramp. The ramps
were linearised to correct for non-linearities in the CRE (Cold-Read-Out
Electronics) output voltages and debiasing effects, which mainly affect the
long wavelength detectors. We attempted to remove glitches resulting from the
impact of high energy particles by applying the two-threshold median filtering
technique, using distributions of 32 data points, three iterations, and
thresholds of three standard deviations for flagging and re-accepting. The
ramps were subdivided into four sections, so as to be more selective in
discarding erroneous read-outs. The first 50\% of the signals per chopper
plateau were discarded to enable the detector response to stabilise at the
level of the source flux density. The chopper plateaus were then treated
separately in deglitching the signals. This was done by running a bin with a
width of 8 signals over the chopper plateau, taking individual steps, and
iterating twice: after maximum/minimum clipping, signals were discarded if
they were off by more than three standard deviations. A correction was applied
to account for the signal dependence on the reset interval which was used for
the read-out sampling. The expected dark current for the orbital position of
the spacecraft was subtracted from the data. The differences between on-source
and interpolated background signals were corrected for signal losses due to
the rapid chopping.

The main source of concern is that the detector response is not constant over
time, and depends on the history of the detector illumination and on the
levels of source and background. Therefore, careful modeling of the detector
behaviour is needed to determine the true source-background signal. At present
the quality of our data and the degree of understanding of the detector
characteristics is insufficient to allow for such an advanced data reduction.
We opted instead for a simpler approach, that also conforms better with the
way the standard star observations and flux-density-calibrations were done.

The in-band power was calibrated using the responsivity as derived from an
FCS1 internal calibrator measurement reduced in the same way as the scientific
measurement, except that background subtraction and chopper frequency
correction are not applicable. The ratio of the responsivities determined from
FCS1 and default varies between 0.8 and 2.8, averaging 1.4 around orbit 200
and 2.2 around orbit 700. We therefore feel that the FCS1 values reveal the
complex detector-history dependence of the responsivity, and hence should be
applied to calibrate the in-band power, instead of using the recommended
default responsivity for chopped observations.

Finally the median flux density was extracted, correcting for the point source
flux density outside of the aperture. The 1-$\sigma$ errors are estimated by
quadratic summation of the flux density uncertainty and the error in the
responsivity value. The 25 $\mu$m flux densities of IRAS05113$-$6739,
IRAS05295$-$7121 and TRM45 have almost certainly been under-estimated due to
off-centering of the source with respect to the ISOPHOT aperture (see the
previous subsection about the CAM imaging).

\subsection{PHOT-C 60 $\mu$m chopped measurements}

Observations at a wavelength of 60 $\mu$m are severely hampered by the complex
and bright sky background compared to the flux density of a typical point
source in our sample. The sky in the direction of the LMC varies on a spatial
scale of a few arcminutes, due to the presence of molecular cloud complexes
within the LMC which are much brighter than circumstellar envelopes at 60
$\mu$m. This scale is not much larger than the PSF for a telescope under 1 m
diameter.

The measurements each consist of 4 DRs (8 for WOH G64), with the number of
NDRs per ramp depending on the expected flux density. Usually we chose 15, 31
or 63 NDRs per ramp, but for IRAS04530$-$6916 and 05329$-$6708 and for WOH G64
only 7 and 3, respectively, were chosen. All DRs were discarded, as well as
the first 3 or 6 NDRs per ramp in case of 31 or 63 NDRs per ramp, respectively
(otherwise only one NDR per ramp was discarded). The ramps were linearised. If
a sufficient number of data points were available, glitches were removed by
applying the two-threshold median filtering technique, using distributions of
16 data points, three iterations, and thresholds of three standard deviations
for flagging and re-accepting. The ramps were subdivided into four sections,
provided there were at least four data points per subramp. The first 50\% of
the signals per chopper plateau were discarded. The signals were deglitched by
running a bin with a width of 16 signals over the chopper plateau, taking
individual steps, and iterating twice: after maximum/minimum clipping, signals
were discarded if they were off by more than three standard deviations. A
reset-interval correction was applied. The expected dark current for the
orbital position of the spacecraft was subtracted from the data. A correction
for vignetting of the $3\times3$ pixels C100 array was applied. The
differences between on-source and interpolated background signals were
corrected for signal losses due to the rapid chopping. This correction is very
large --- a factor of 2.75 --- and rather uncertain (35\%).

The in-band power was calibrated using the responsivity as derived from an
FCS1 internal calibrator measurement done immediately after the scientific
measurement. The FCS responsivities were found to resemble the pixel response
pattern in the scientific measurement better than the default responsivities
did. The FCS measurement was reduced in the same way as the scientific
measurement, except that vignetting correction, background subtraction and
hence a chopper frequency correction are not applicable. The ratio of the FCS1
and default responsivities varied between 0.8 and 2.0, around a median of 1.3
and with a standard deviation of 0.3. It varied from orbit to orbit and within
an orbit, but with no clear time dependence --- except that it was larger
after orbit 190 than before. This behaviour closely resembles that seen for
the PHOT-P detectors.

Finally the median flux densities of the pixels were extracted. The median
value of the eight pixels surrounding the central one was subtracted from the
central pixel value. Knowing that the central pixel contains 66.35\% of the
total flux density of a well-centred point source, and the entire C100 array
contains 91.75\%, a correction factor of 1.5835 was applied. The (internal)
1-$\sigma$ errors are estimated by quadratic summation of the error in the
value for the central pixel and $1/\sqrt{8}$ times the median of the
differences between background pixels and their median. This does not take
into account the uncertainty in the absolute flux-density calibration. The
flux density of IRAS05295$-$7121 has almost certainly been under-estimated due
to off-centering of the source with respect to the ISOPHOT aperture (see the
earlier subsection about the CAM imaging).

\subsection{PHOT-C 60 $\mu$m mapping observations}

The measurements each consist of 64 DRs, with 63 NDRs per ramp (128 DRs and 31
NDRs per ramp for IRAS05402$-$6956 and WBP14). All DRs were discarded, as well
as the first 3 or 6 NDRs per ramp in case of 31 or 63 NDRs per ramp,
respectively. The ramps were linearised. Glitches were removed by applying the
two-threshold median filtering technique, using distributions of 32 data
points, three iterations, and thresholds of three standard deviations for
flagging and re-accepting. The ramps were subdivided into four sections. The
first 4 read-outs per raster point were discarded. The signals were deglitched
by running a bin with a width of 64 signals through the raster point, taking
individual steps, and iterating twice: after maximum/minimum clipping, signals
were discarded when they were off by more than three standard deviations. We
applied the stability recognition method to derive the signal most likely to
be near the true, stabilised one: a bin with a width of 64 signals was run
over the measurement, with intervals of 32 signals, and only data within a
confidence level of 0.95 were kept. A reset-interval correction was applied.
The expected dark current for the orbital position of the spacecraft was
subtracted from the data. The $3\times3$ pixels array was corrected for
vignetting.

The in-band power was calibrated by interpolating the responsivities derived
from FCS1 internal calibrator measurements made before and after the
scientific measurement. Use of the two FCS measurements was found to correct
(in part) for drift behaviour of the signal, which was not the case if default
responses were used. The FCS measurements were reduced in the same way as the
scientific measurement, except that a vignetting correction is not applicable
and 16 data points were used for signal deglitching and stability recognition.
Finally the median flux densities of the pixels were extracted.

The values of the eight pixels surrounding the central one were replaced by
their median value and added to the central pixel value. The values of the
sixteen pixels along the rim of the map were also replaced by their median
value, and this was taken to be the sky background to be used for correcting
the total flux density in the inner $3\times3$ part of the map. Knowing that
the central pixel contains 66.35\% of the total flux density of a well-centred
point source, and the entire C100 array contains 91.75\%, the outer sixteen
pixels contain a fraction of the flux density that is somewhere between 0 and
8.25\%. Assuming this fraction to be 4.125\%, a factor of 1.1182 is derived to
correct the background subtracted flux density for the stellar flux density
contained in the outer sixteen pixels. The (internal) 1-$\sigma$ errors are
estimated by quadratic summation of the error in the value for the central
pixel, $8/\sqrt{8}$ times the median of the differences between the eight
surrounding pixels and their median, and $9/\sqrt{16}$ times the median of the
differences between the outer sixteen pixels and their median. The error in
the value for the central pixel was determined by the median of the
differences between the values for the pixels that were centred on the star
over the course of the mapping, and their median. The error estimate does not
take into account the uncertainty in the absolute flux-density calibration.

\section{ISO spectroscopy}

\subsection{CAM-CVF spectro-photometry}

The CAM-CVF cube $(X,Y,\lambda)$ was corrected for the dark image using a
model for the dependence on the revolution and orbital position of the
spacecraft, and corrected for glitches by applying a multiresolution median
transform. For each pixel, we applied a $\kappa$-$\sigma$ rejection criterion
to the evolution of the signal in time: values that deviate more than 2
$\sigma$ from the time-average were rejected. Attempts to correct for the
transient behaviour of the signal by applying various models yielded results
no better than this method. The images were divided by the default calibration
flatfield image to correct for the array pattern in the pixel responsivities,
and time-averaged.

The spectra were constructed by obtaining photometry from the images that each
correspond to a different position of the CVF. For each target we integrated
the signal over the pixels that were significantly above the background level.
The number of pixels was limited to avoid sampling excessive background, and
ranged from 1, for the faintest, to 9, for the brightest sources. The short-
and long-wavelength parts of the CVF were treated separately because the star
was often centred at a slightly different position on the array. The background
level was determined by taking the median value within a three-pixel wide
circular annulus around the star, with an inner radius of three pixels. The
stellar flux density was corrected for the wavelength dependence of the PSF.
This correction was derived by doing photometry on the PSF with equivalent
integration areas but inversely proportional to the wavelength. The first few
($\sim4$) spectro-photometric points are prone to have flux densities that are
over-estimated due to stabilisation difficulties when the CVF scanned from the
shortest towards longer wavelengths.

\subsection{PHOT-S spectro-photometry}

The measurements each consist of 4 or 8 DRs, with 127 NDRs per ramp, except
for WOH G64 (16 DRs and 63 NDRs per ramp). All DRs were discarded, as well as
the first 12 NDRs per ramp (6 in case of WOH G64). Glitches were removed by
applying the two-threshold median filtering technique, using distributions of
32 data points, three iterations, and thresholds of three standard deviations
for flagging and re-accepting. The ramps were subdivided into four sections.
The first four signals per measurement were discarded. The signals were
deglitched by running a bin with a width of 16 signals over the chopper
plateau, taking individual steps, and iterating twice: after maximum/minimum
clipping, signals were discarded when they were off by more than three
standard deviations. The stability recognition method was applied, using an
8-signals wide bin and intervals of 4 signals, keeping only data within a
confidence level of 0.95. The time evolution of the signal for each pixel was
checked by eye, and obviously bad data that had passed the earlier rejection
criteria were removed by hand at this stage. The expected dark current for the
orbital position of the spacecraft was subtracted from the data.

The in-band power was calibrated using the default responsivities expected for
the orbital position, and finally the mean flux density was extracted, taking
into account the amount of flux density lost outside of the aperture. Note
that the wavelength dependence of the responsivity shows a bump around 3
$\mu$m. When applied to the spectrum of a star, this could introduce an
artificial spectral feature which might be interpreted erroneously as
absorption due, for instance, to H$_2$O ice.

%
% FIGURE B1
%
\begin{figure}[tb]
\centerline{\psfig{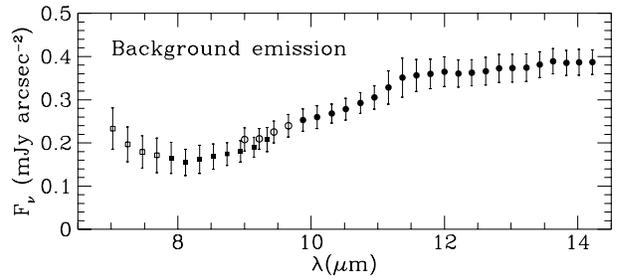}}
\caption[]{The CAM-CVF background emission spectrum. Open symbols represent
spectro-photometric points that are prone to have flux densities that are
over-estimated due to stabilisation difficulties. The spectral shape is best
represented by the solid symbols (squares for the short-, disks for the
long-wavelength region). The background emission actually arises from the
zodiacal dust belt in our Solar system.}
\end{figure}

The PHOT-S observations did not include a separate measurement of the
background. We used the average of the background spectra measured in the
CAM-CVF images (Fig.\ B1, errobars indicate the standard deviation in the set
of 12 spectra). This spectrum was used for wavelengths $>8$ $\mu$m, and
interpolated linearly below 8 $\mu$m to a value of zero at 5 $\mu$m (at
shorter wavelengths the background is assumed to be negligible). We ignore the
first 4 steps of the CVF filter wheel, that are usually not stabilised. The
background is predominantly emission from the zodiacal dust belt in our
Solar System. Because of the annual modulation of the zodiacal light level in
a particular direction of the sky, the weekly all-sky maps produced by the
DIRBE instrument onboard the COsmic Background Explorer (COBE)
(http://www.gsfc.nasa.gov/astro/cobe/\#dirbe) were used to scale the
background spectrum to the epochs of the PHOT-S spectra (see \'{A}brah\'{a}m
et al.\ 1998). The COBE/DIRBE 12 $\mu$m surface brightnesses for our CVF and
PHOT-S observations have standard deviations of 5.5 and 5.9\%, respectively.
The CVF background spectra yield a standard deviation at 12 $\mu$m of 9.1\%.
Although the CVF and COBE/DIRBE do correlate, the spread in the CVF background
exceeds the variation in the zodiacal light, suggesting that variations in the
background emission from the LMC is discernible in our data. The scaled
background spectrum is subtracted from the PHOT-S spectra, that were obtained
through a $24^{\prime\prime}\times24^{\prime\prime}$ aperture.

IRAS05112$-$6755 was observed twice. The spectra look identical, except that
the second is a factor 1.5 brighter than the first, independent of wavelength,
whilst the near-IR flux densities only differed by $\sim$10\%. Pointing
thresholds were $10^{\prime\prime}$ and $2^{\prime\prime}$, respectively, and
the star may have been close to the edge of the aperture during the first
observation. Flux density levels in the PHOT-S spectrum of IRAS05298$-$6957
may have been under-estimated due to off-centering of the star in the PHOT-S
aperture. The PHOT-S spectrum of IRAS04496$-$6958 is fainter than the CAM-CVF
spectrum, whereas for IRAS05558$-$7000 the reverse is true. This may be a
result of variability, as seen from the near-IR photometry at the epochs of
these ISO observations.

\end{document}